\documentclass{aa}  

\usepackage{graphicx}
\usepackage{txfonts}
\newcommand{\dg}{$^{\circ}$}
\newcommand{\al}{$\alpha$}
\newcommand{\kms}{km.s$^{-1}$}
\usepackage{color}

\begin{document}

   \title{Stability of the interstellar hydrogen inflow longitude from 20 years of SOHO/SWAN observations}
   \titlerunning{Interstellar hydrogen inflow longitude from 20 years of SOHO/SWAN data}
   \subtitle{}

   \author{D. Koutroumpa\inst{1}%\fnmsep\thanks{Just to show the usage
%          of the elements in the author field}
          \and
          E. Qu\'emerais\inst{1}
          \and
          O. Katushkina\inst{2}
          \and
          R. Lallement\inst{3}
          \and
          J.-L. Bertaux\inst{1}
          \and
          W. Schmidt\inst{4}
         }

   \institute{Universit\'e Versailles St.-Quentin; Sorbonne Universit\'es, UPMC Univ. Paris 06; CRNS/INSU, LATMOS-IPSL, 11 boulevard d'Alembert, 78280 Guyancourt, France\\
   \email{dimitra.koutroumpa@latmos.ipsl.fr}
         \and
         Space Research Institute of Russian Academy of Sciences, Moscow, 117997, Russia
         \and
         GEPI/Observatoire de Paris, 5 Place Jules Janssen, 92195 Meudon, France
         \and
         Finnish Meteorological Institute, Helsinki, Finland\\             
%             \thanks{}
             }

   \date{Received September 15, 1996; accepted March 16, 1997}

% \abstract{}{}{}{}{} 
% 5 {} token are mandatory
 
  \abstract
   {}
   {A recent debate on the decade-long stability of the interstellar He flow vector, and in particular the flow longitude, has prompted us to check for any variability in the interstellar H flow vector as observed by the SWAN instrument on board SOHO.}
   {We used a simple model-independent method to determine the interstellar H flow longitude, based on the parallax effects induced on the Lyman-\al\ intensity measured by SWAN following the satellite motion around the Sun.}
   {Our results show that the interstellar H flow vector longitude does not vary significantly from an average value of 252.9\dg\ $\pm$ 1.4\dg\ throughout the 20-year span of the SWAN dataset, further strengthening the arguments for the stability of the interstellar gas flow.}
   {}


   \maketitle
%
%-------------------------------------------------------------------

\section{Introduction}

The Sun moves through the diffuse partially ionized interstellar gas at a relative velocity of $\sim$25 \kms. The supersonic solar wind (SW) carves out a `bubble', the heliosphere, through the interstellar plasma, confining the SW plasma inside the heliosphere. On the outside, the interstellar plasma is forced to heat, decelerate, and deviate around the heliospheric interface \citep[e.g.,][]{Baranov1993}. The neutral particles (principally hydrogen with about 15\%\ helium and a small portion of oxygen atoms) of the interstellar medium have mean free paths of the scale of the heliospheric interface \citep[$\sim$ 100 AU, ][]{Baranov1993} and can therefore penetrate through this region into the solar system in the form of an interstellar wind in the rest frame of the Sun.

The H population is filtered at the heliospheric interface through resonant charge-exchange coupling with protons \citep[e.g.,][]{Izmodenov1998,Izmodenov2007}. This process creates a population of secondary neutral atoms that conserve the local dynamic properties of the shocked interstellar plasma (heating, deceleration, and deflected velocity vector) at the heliospheric interface. A mixture of the two neutral populations (primary and secondary) then propagates into the interplanetary medium and can be observed near the Earth's orbit via scattered solar Lyman-\al\ radiation, for example. The Solar Wind ANisotropies (SWAN) experiment \citep{Bertaux1995} on board the Solar and Heliospheric Observatory (SOHO) has been mapping the backscattered Lyman-\al\ radiation in the interplanetary medium for more than 20 years and allowed a firm determination of the interstellar H flow vector coordinates ($\lambda$, $\beta$) = (252.5\dg $\pm$ 0.7\dg, 8.9\dg $\pm$ 0.5\dg) thanks to Doppler-shift analyses of the backscattered resonance line using the SWAN hydrogen absorption cell \citep[combined values from][]{Quemerais1999,Lallement2005,Lallement2010a}. A recent analysis, using SOHO/SWAN and MESSENGER/MASCS (Mercury Atmospheric and Surface Composition Spectrometer) mutual observations during the MESSENGER cruise to Mercury in 2010 and 2011 and a triangulation method of the position of the maximum emissivity region (MER), confirmed the H flow vector longitude at 253.2\dg\ $\pm$ 2\dg\ \citep{Quemerais2014}.

The He cross-section for charge-exchange is much smaller, and therefore the population that enters the heliosphere preserves the characteristics of the local interstellar medium. Following a compilation of several UV, pick-up ion, and neutral particle measurements, a coordinated effort by an international team gathered at the International Space Science Institute (ISSI) in 2003-2004 solved a long-standing systematic discrepancy between in situ and particle measurements \citep{Lallement2004c} and defined a weighted mean set of parameters that best reproduced all the datasets available at the time. These parameters were ($\lambda$, $\beta$) = (74.7\dg, -5.3\dg) downwind, translated into (254.7\dg, 5.3\dg) upwind, V = 26.2 \kms, T = 6300 K, and n = 0.015 cm$^{-3}$ \citep[see ][for a synopsis]{Mobius2004}. The difference between H and He incoming directions has been explained by the perturbation of H flow around the heliospheric interface, allowing an estimate of the direction of the interstellar magnetic field \citep{Lallement2005}.

The recent analyses of neutral atom measurements with the Interstellar Boundary Explorer low energy channel (IBEX-Lo) \citep{Mobius2012, Bzowski2012} have deduced quite different parameters for the interstellar He flow (ecliptic longitude $\sim$ 259\dg, ecliptic latitude $\sim$5\dg\ , velocity $\sim$23 \kms, and a broad temperature range between 5000 K and ~8000 K) from what is traditionally agreed upon. Coincidentally, the difference was impressive for the flow longitude (over 4\dg\ in ecliptic longitude) and velocity, while the latitude remained basically unchanged. Another analysis \citep{Frisch2013} took the matter a step further, claiming that the longitude difference shows a decade-long trend, based on a compilation of various data samples since the 1970s, although the validity of this analysis was strongly challenged by \citet{Lallement2014}, who also pointed out an additional suspicious coincidence in the parameter set. Two more recent papers have also argued against the new IBEX values. The first \citep{Katushkina2014} reanalyzed the Ulysses/GAS data and IBEX-Lo data using a time-dependent 3D kinetic model, showing that the IBEX-found flow vector is incompatible with the Ulysses/GAS data, while the standard ISSI team consensus of vector parameters might in principle fit the IBEX-Lo data if the gas temperature were increased from 6300 K to 9000 K. The second paper \citep{Vincent2014} fitted HST/STIS data of the interplanetary H Lyman-\al\ line during solar cycle 24 and compared the results (along with a compilation of SWAN data up to 2003) to realistic 3D gas-dynamic models to show that the interstellar flow velocity vector has not changed significantly since the Ulysses measurements. More recently even, a refined analysis of the first six years of IBEX data has also recanted to the standard consensus vector for He, but with a much higher temperature (7000-9500 K) to reconcile the IBEX and Ulysses data \citep{Bzowski2015, McComas2015, Mobius2015, Schwadron2015, Sokol2015, Swaczyna2015}.

%--------------------------------------------------------------------

\begin{figure}
   \centering
   \includegraphics[width=6.5cm]{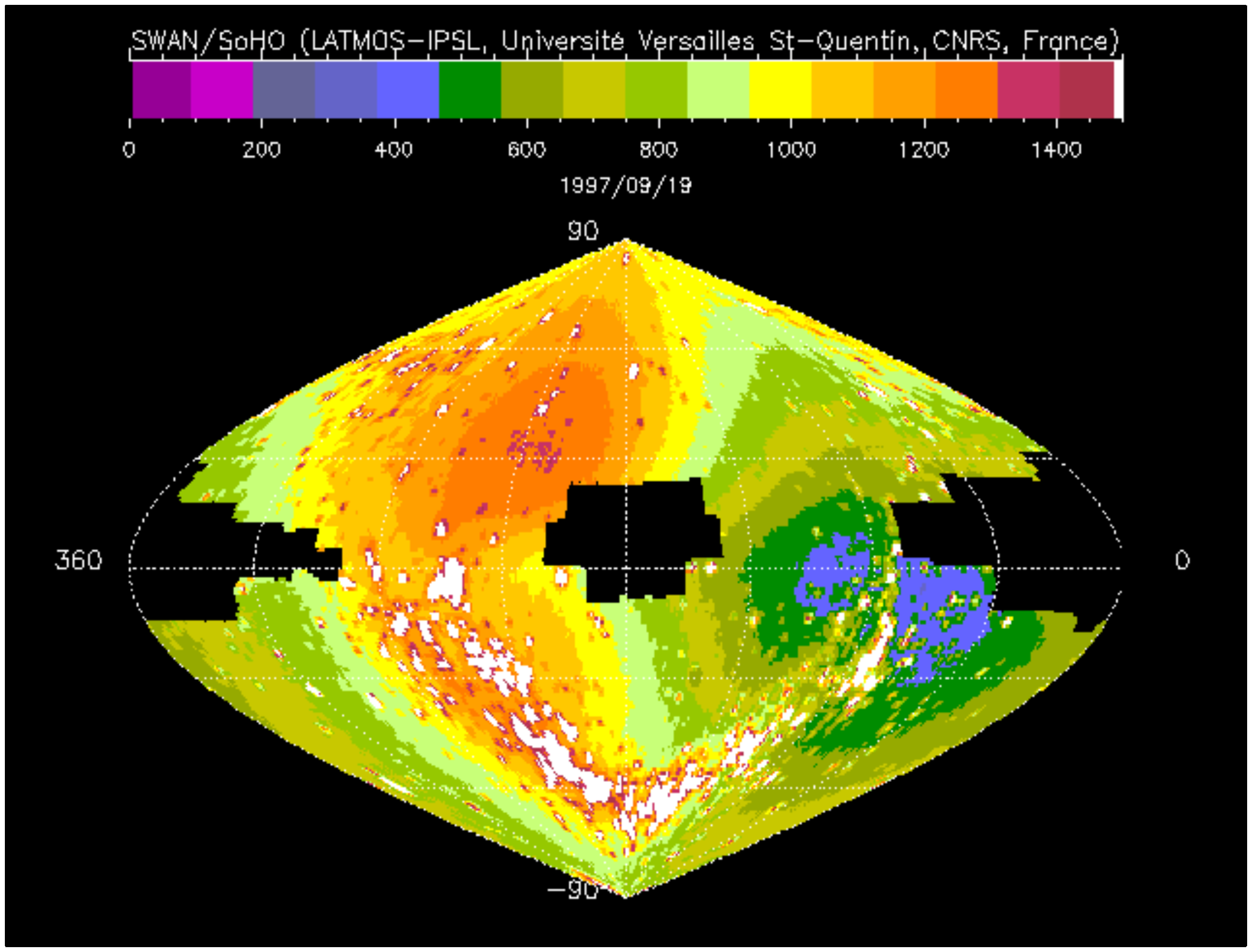}
    \includegraphics[width=6.5cm]{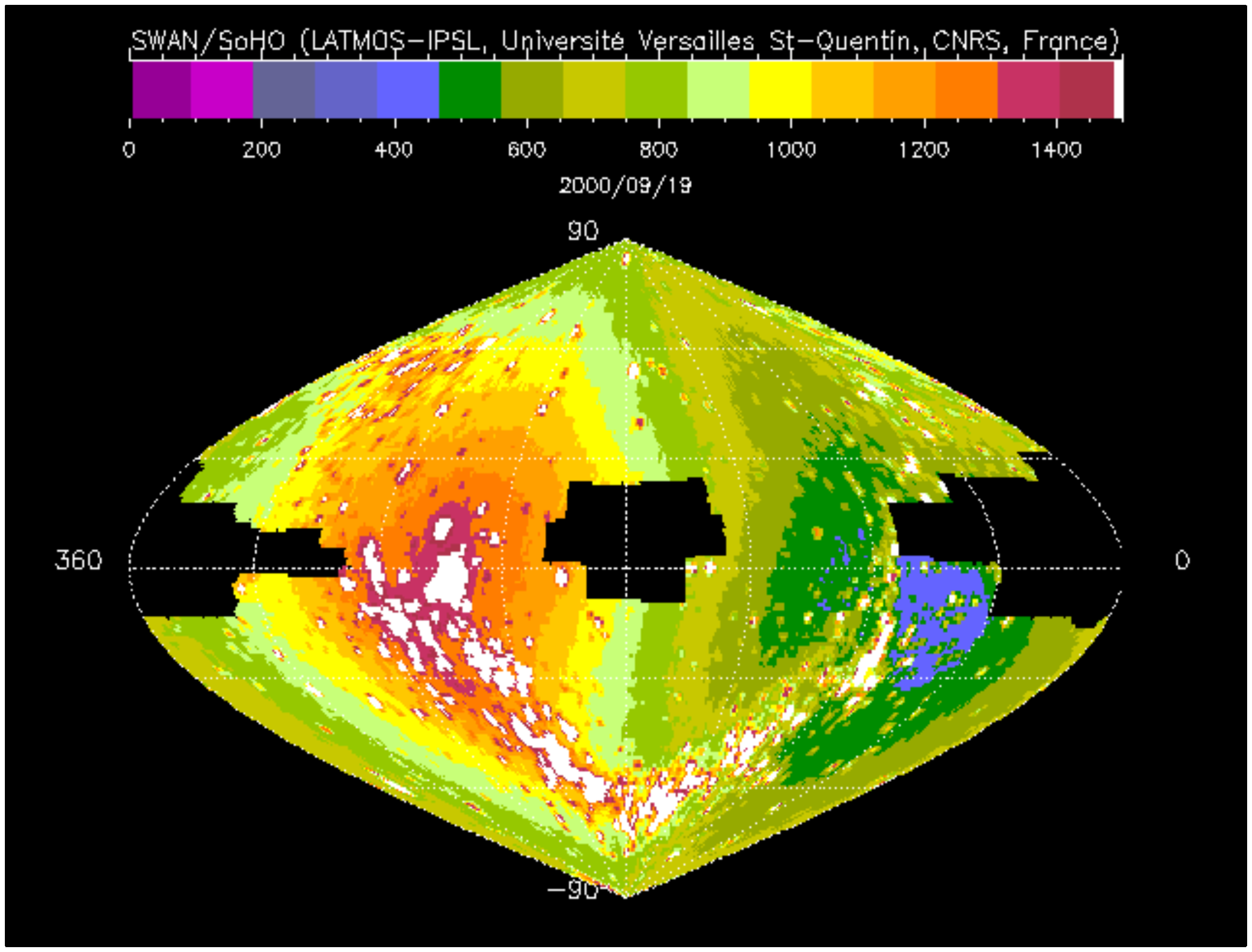}
\caption{Full-sky Lyman-\al\ emission maps in ecliptic (J2000) coordinates recorded by SOHO/SWAN in September 1997 (top panel) and September 2000 (bottom panel). The maps are presented in ecliptic coordinates centered at $\sim$180\dg, with ecliptic longitude increasing from right to left.The color scales are in Rayleighs. The two masked areas correspond to the shadow of the spacecraft (centered at $\sim$0\dg) and to the shadows of the Sun shields that protect the sensors from direct sunlight (centered at $\sim$180\dg). }\label{figMaps}
\end{figure}

%--------------------------------------------------------------------

\begin{figure}
   \centering
   \includegraphics[width=6.5cm]{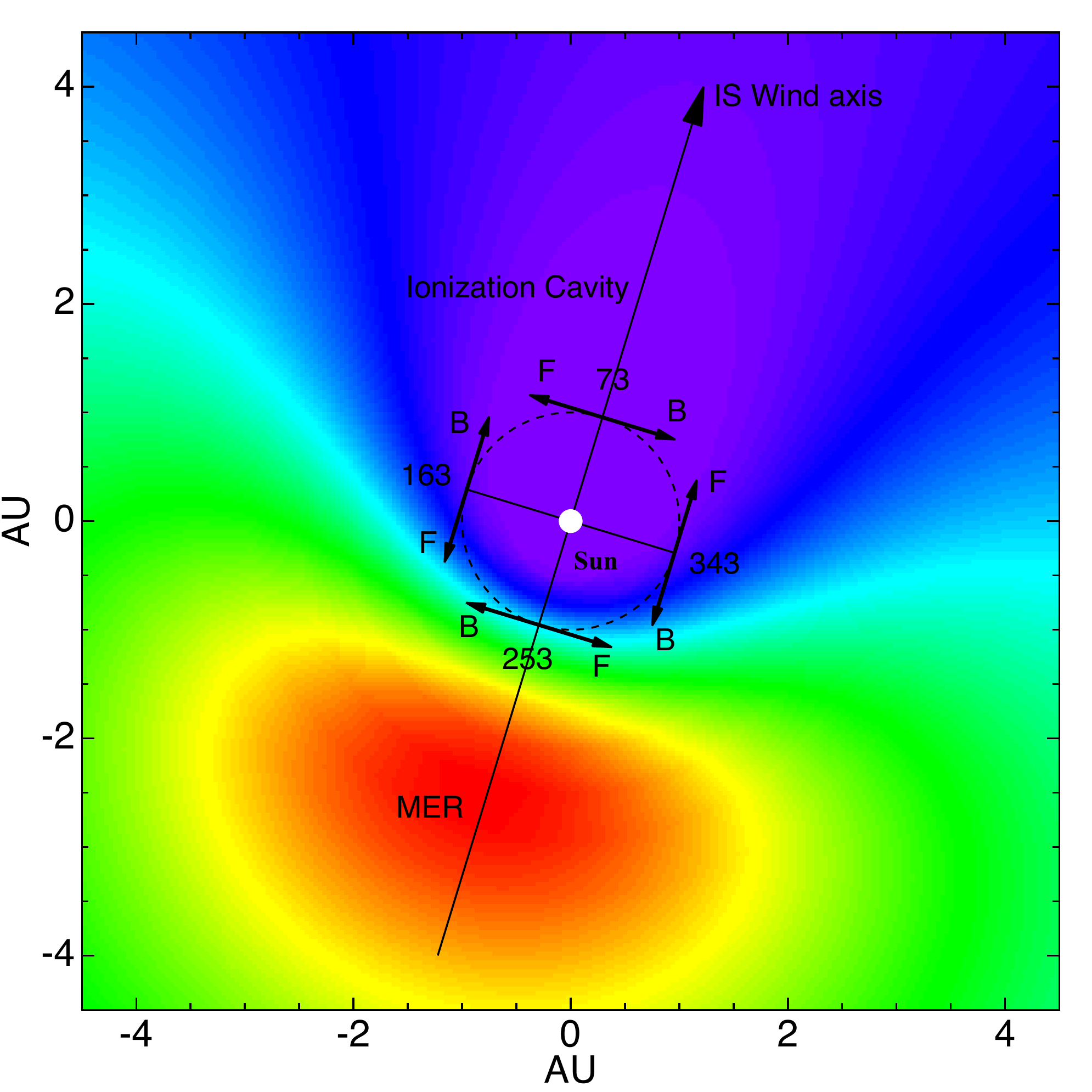}
\caption{Simulated Lyman-$\alpha$ emissivity (color scale is in arbitrary units) within the ecliptic plane. The dashed circle shows the approximate SOHO orbit around the Sun and the small arrows show a sample of the selected lines of sight (marked F for forward and B for backward-looking w.r.t. Earth/SOHO motion) for this analysis, projected onto the ecliptic plane. A projection of the interstellar H flow axis on the ecliptic plane is also marked. } \label{figGeo}
\end{figure}

%-------------------------------------- Two column figure (place early!)
   \begin{figure*}
   \centering
   \includegraphics[width=17cm]{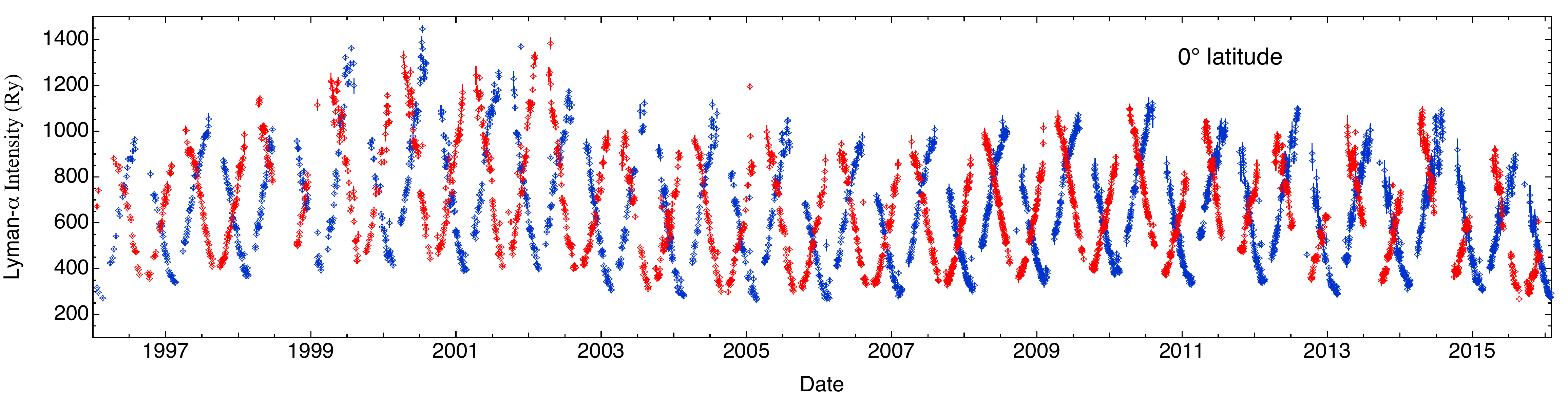}
   \caption{Lyman-$\alpha$ intensities as a function of time for the 0\dg\ latitude coupled lines of sight. Red points represent the forward-pointing (w.r.t. Earth/SOHO motion) line of sight and blue points represent the backward-pointing line of sight. The biannual crossing of the interstellar plane of symmetry where the forward and backward lines of sight point through equal-emitting gas volume is clearly visible. }
               \label{figLya2time}%
    \end{figure*}

Any variation in the He flow should also be reflected in the primary H population, and therefore induce changes in the global average H flow observed near the Sun. In this paper we analyze 20 years of SWAN data to check for variability in the interstellar H flow direction. We present a simple and model-independent method \citep[similar to the triangulation used by][]{Quemerais2014} to determine the H flow axis ecliptic longitude in Sect.~\ref{Sec2} and discuss the stability of the results given the influence of solar activity on the interplanetary H Lyman-\al\ emission in Sect.~\ref{Sec3}.

%--------------------------------------------------------------------

\section{Data selection and analysis}\label{Sec2}
Operating on board the SOHO spacecraft quasi-uninterrupted since January 1996, the SWAN instrument \citep{Bertaux1995} is a UV photometer with a passband between 110 nm and 160 nm. Two identical units placed at opposite sides of the SOHO spacecraft (+Z and -Z sides), each equipped with a mechanical periscopic scanning system, allow it to produce full-sky Lyman-\al\ intensity maps with 1\dg\ by 1\dg\ resolution approximately once a day. Details on the operations of the instrument and data processing for the reconstruction of the Lyman-\al\ maps are given in \citep{Quemerais1999, Quemerais2006a}. Each of the sensors is also equipped with a hydrogen absorption cell, which allows for a fine spectral analysis of the Lyman-\al\ line profile \citep{Quemerais2006}. Here we focus on the photometric data of the instrument.

%-------------------------------------- 
\begin{figure}
   \centering
   \includegraphics[width=8.5cm]{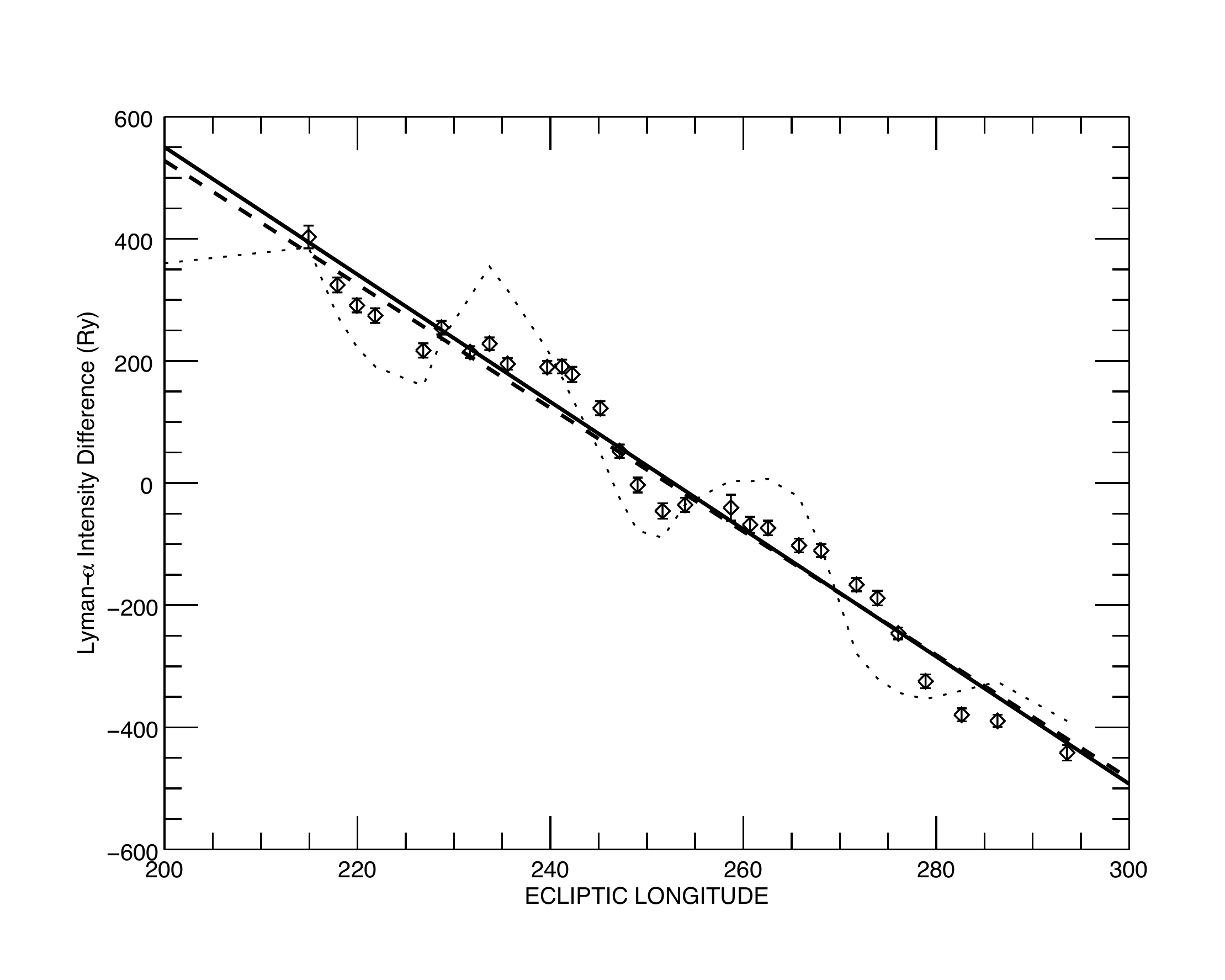}
\caption{Example of SWAN data fit for 1999, at -29\dg\ latitude. The diamonds and full line represent the data points and fit for solar UV-corrected data, respectively. The dotted and dashed lines represent the data and fit for the uncorrected case, respectively.} \label{fig1999Fit}
\end{figure}

%-------------------------------------- Two column figure (place early!)
   \begin{figure*}
   \centering
   \includegraphics[scale=0.2]{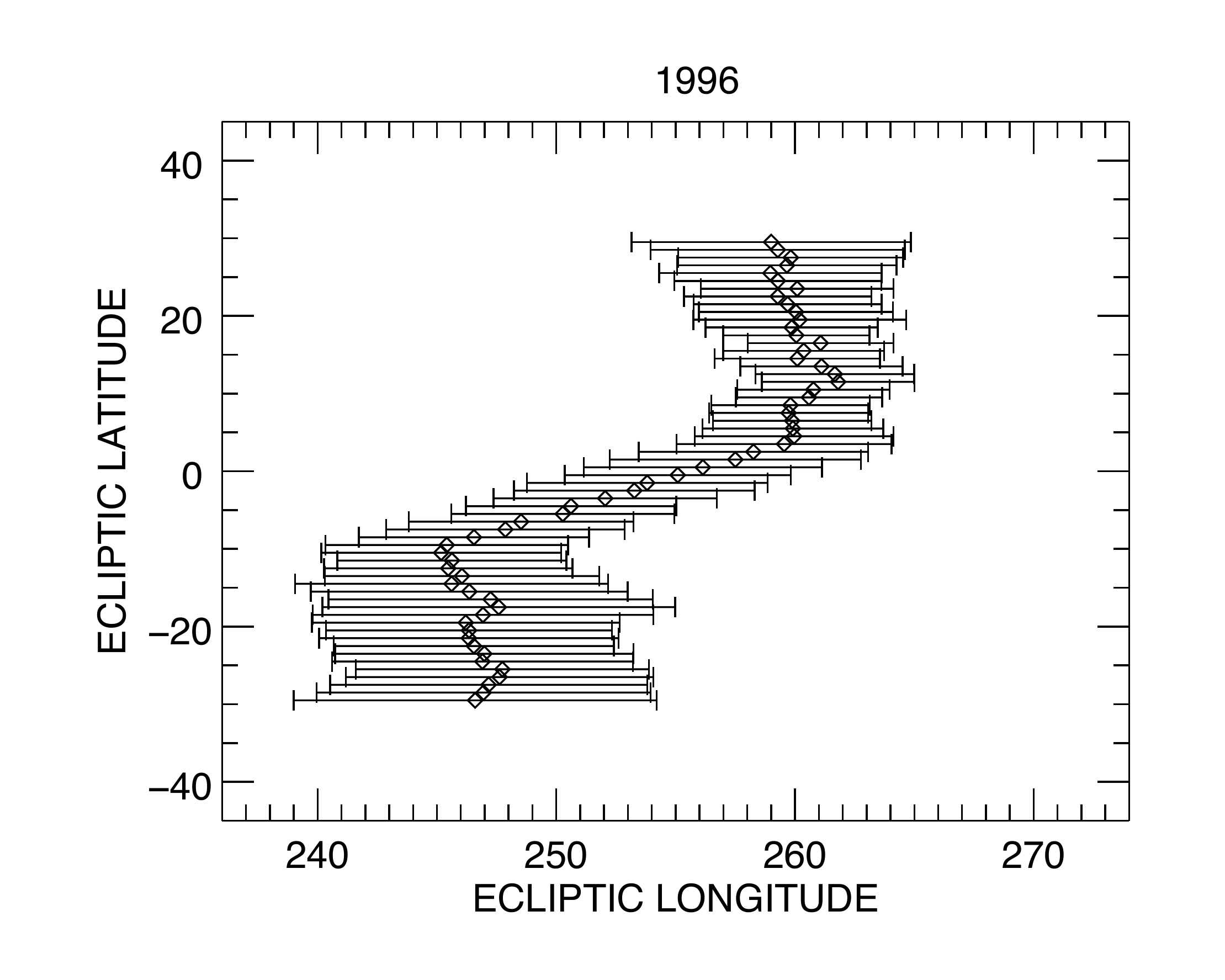}
   \includegraphics[scale=0.2]{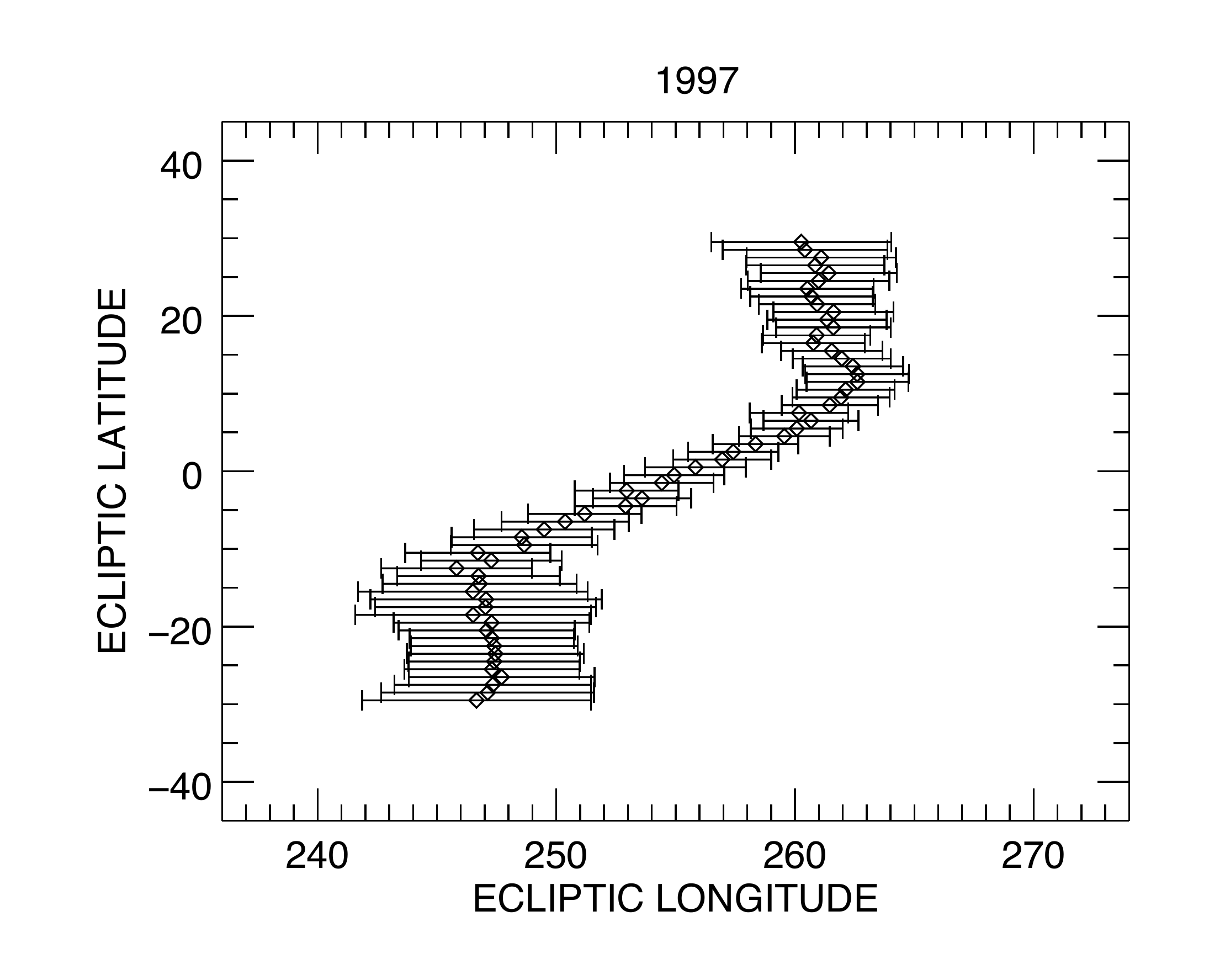}
   \includegraphics[scale=0.2]{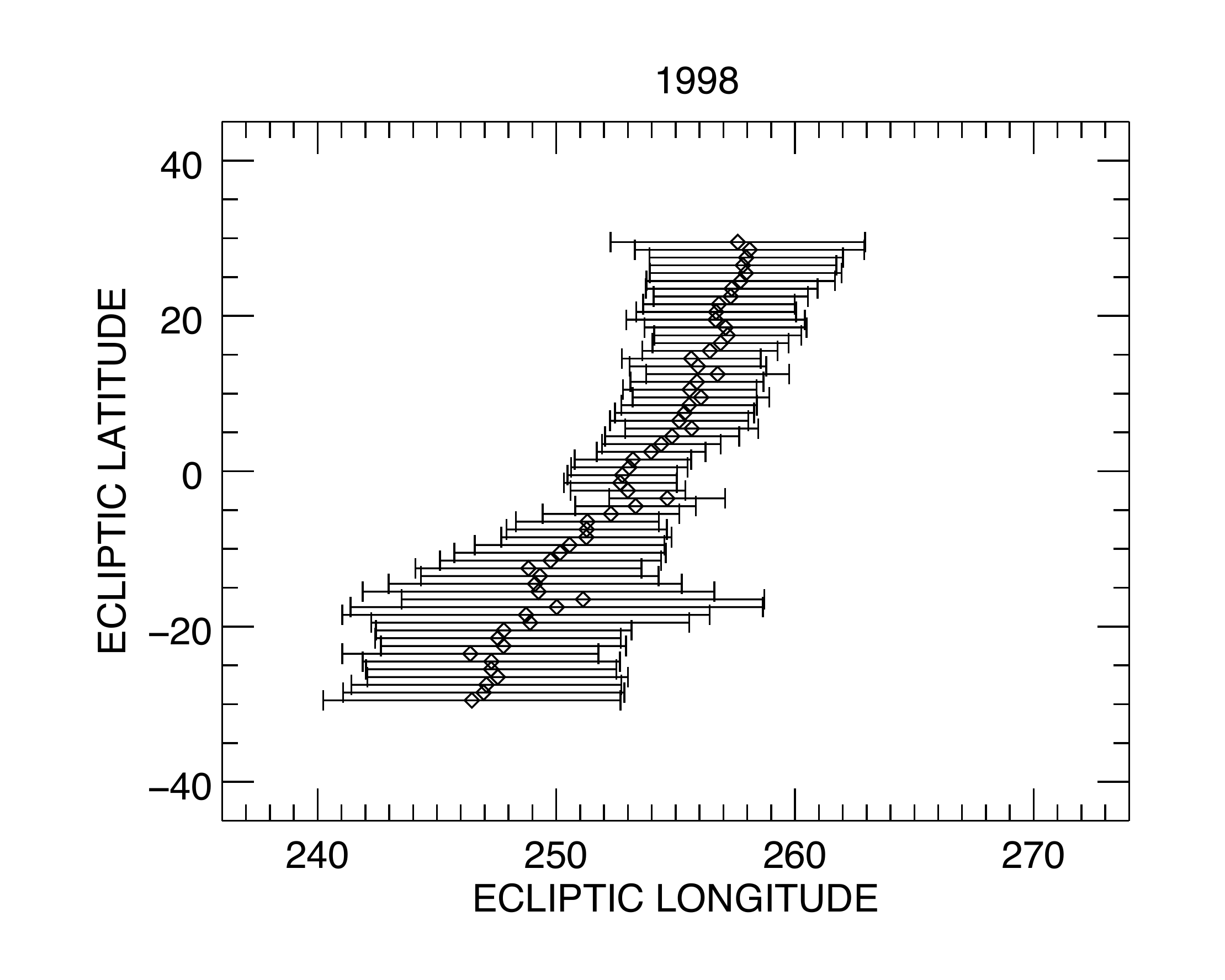}
   \includegraphics[scale=0.2]{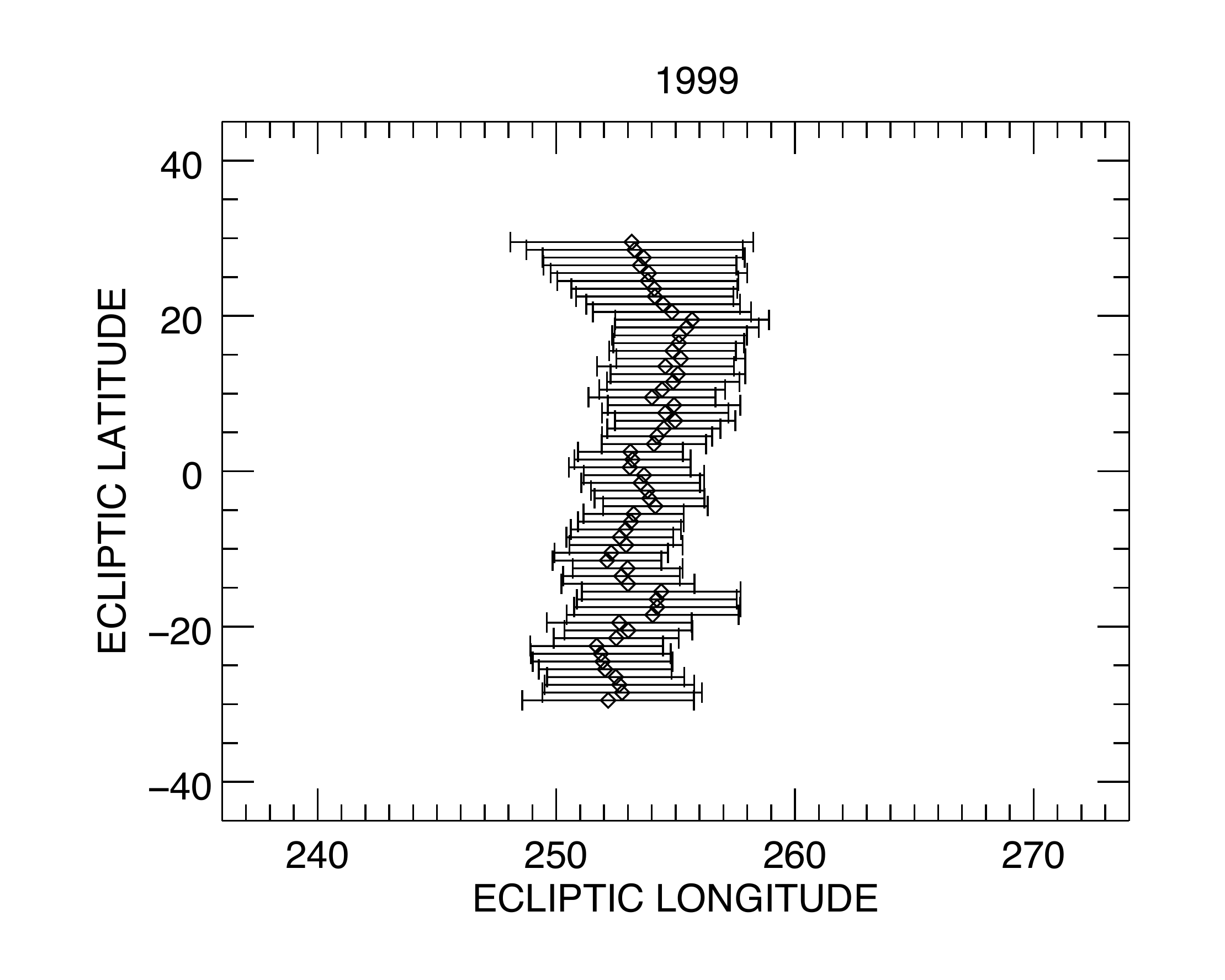}
   \includegraphics[scale=0.2]{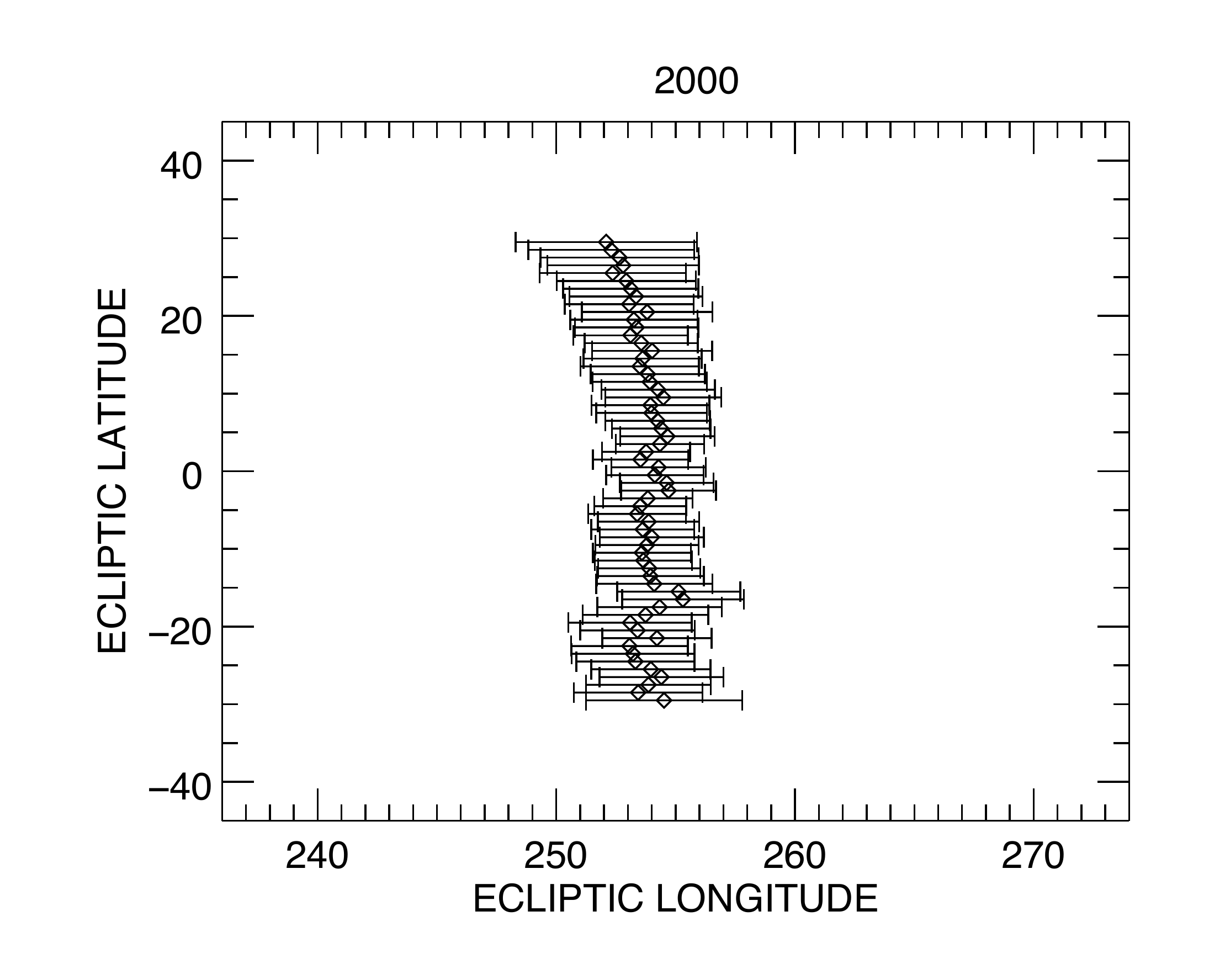}
   \includegraphics[scale=0.2]{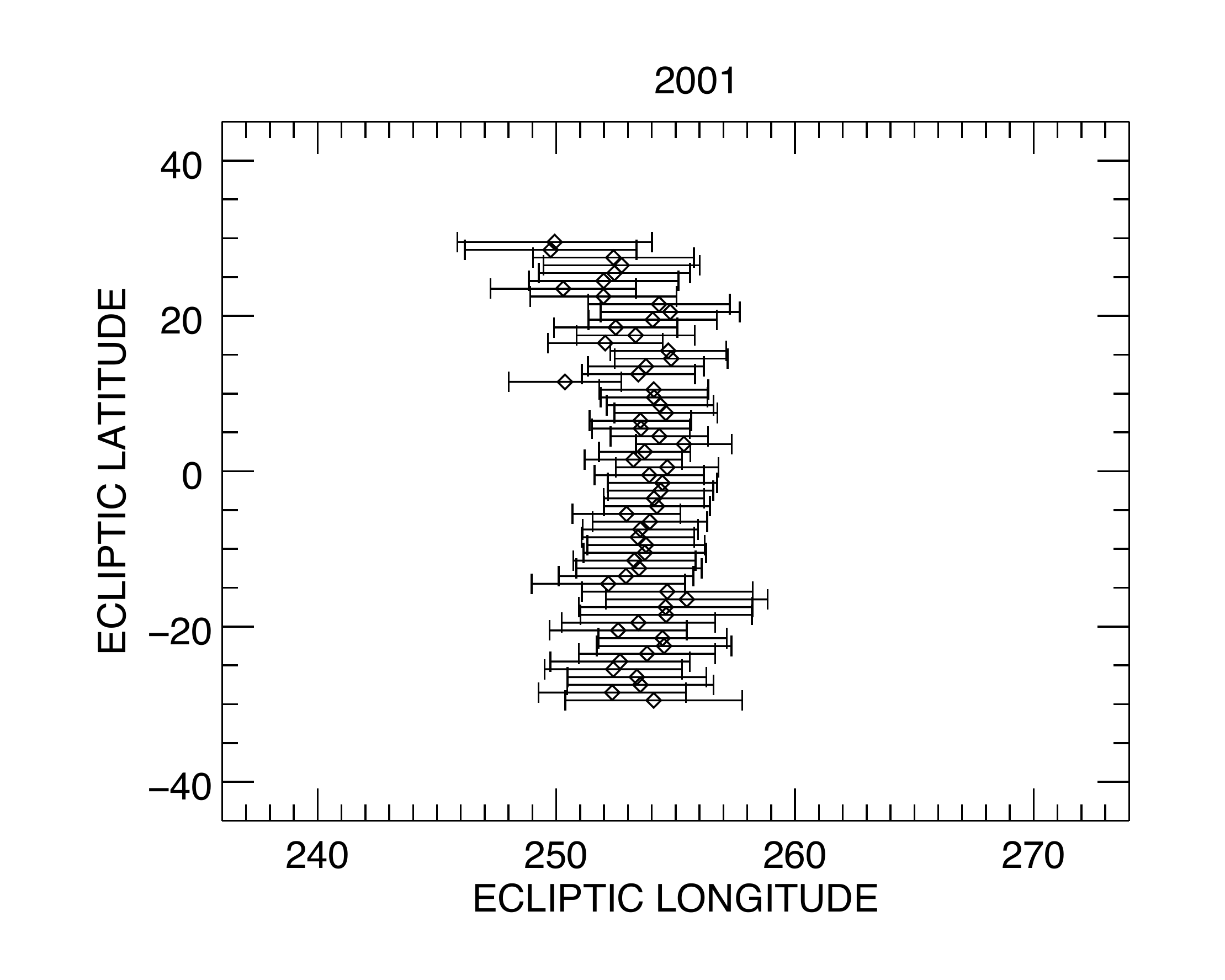}
   \includegraphics[scale=0.2]{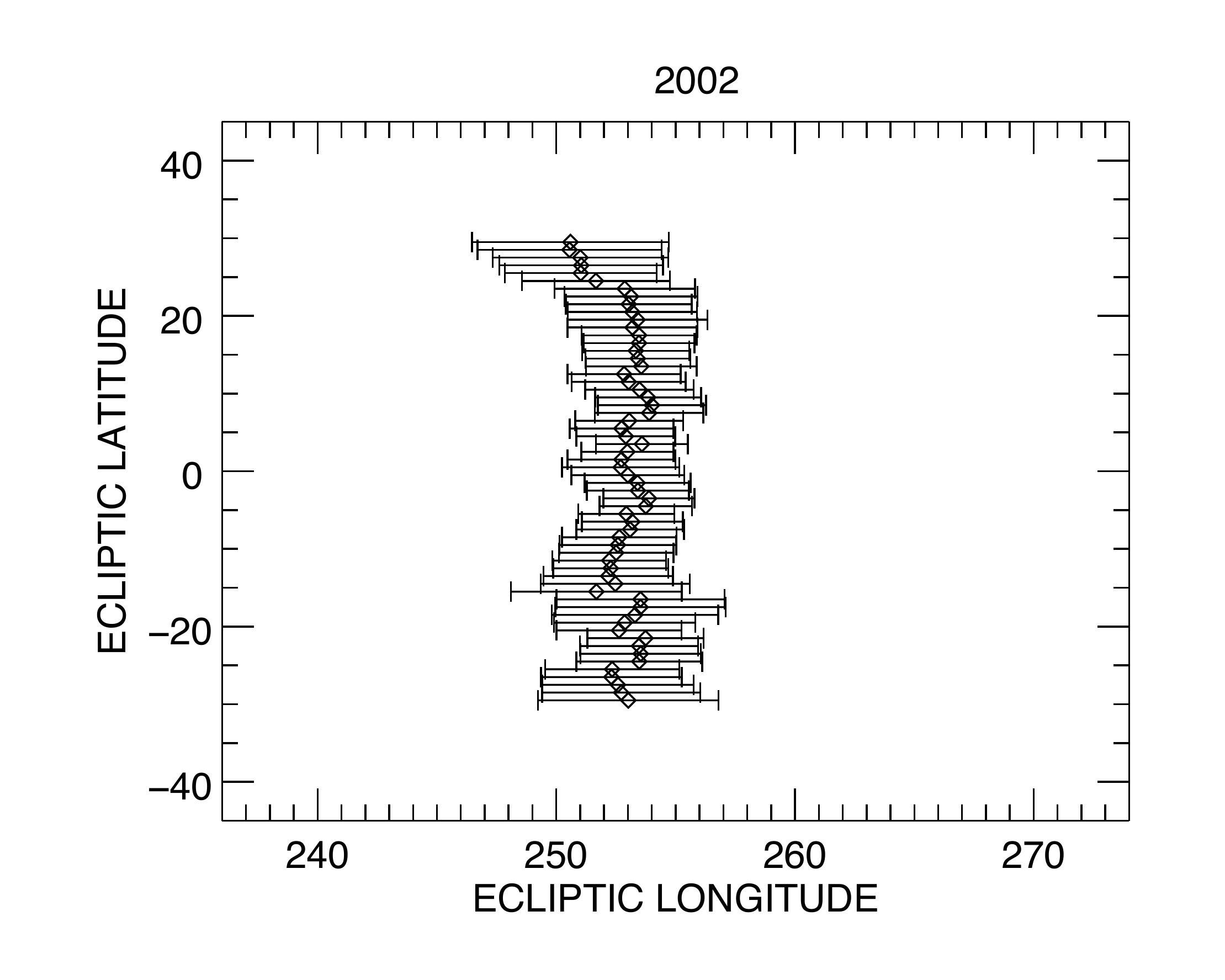}
   \includegraphics[scale=0.2]{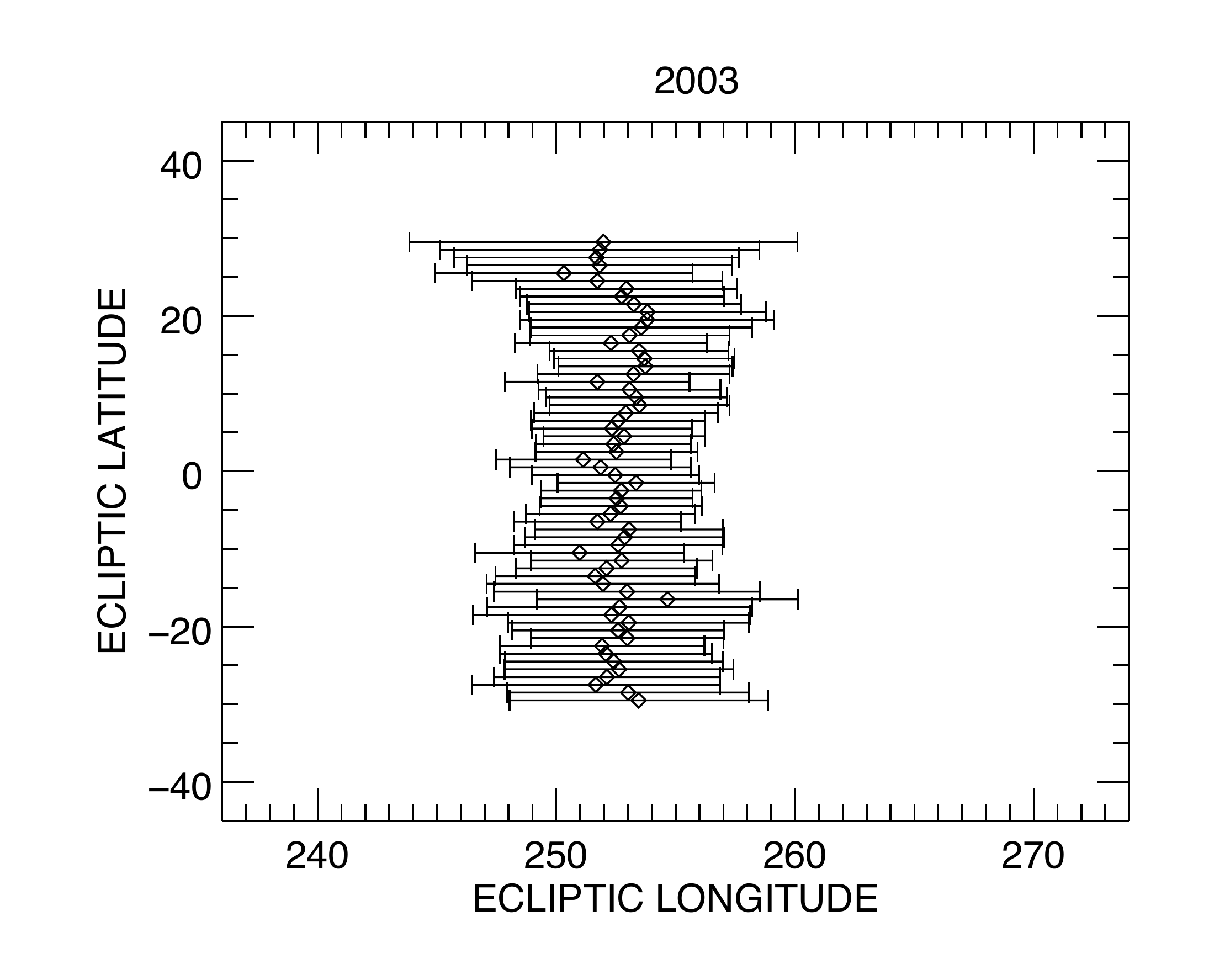}
   \includegraphics[scale=0.2]{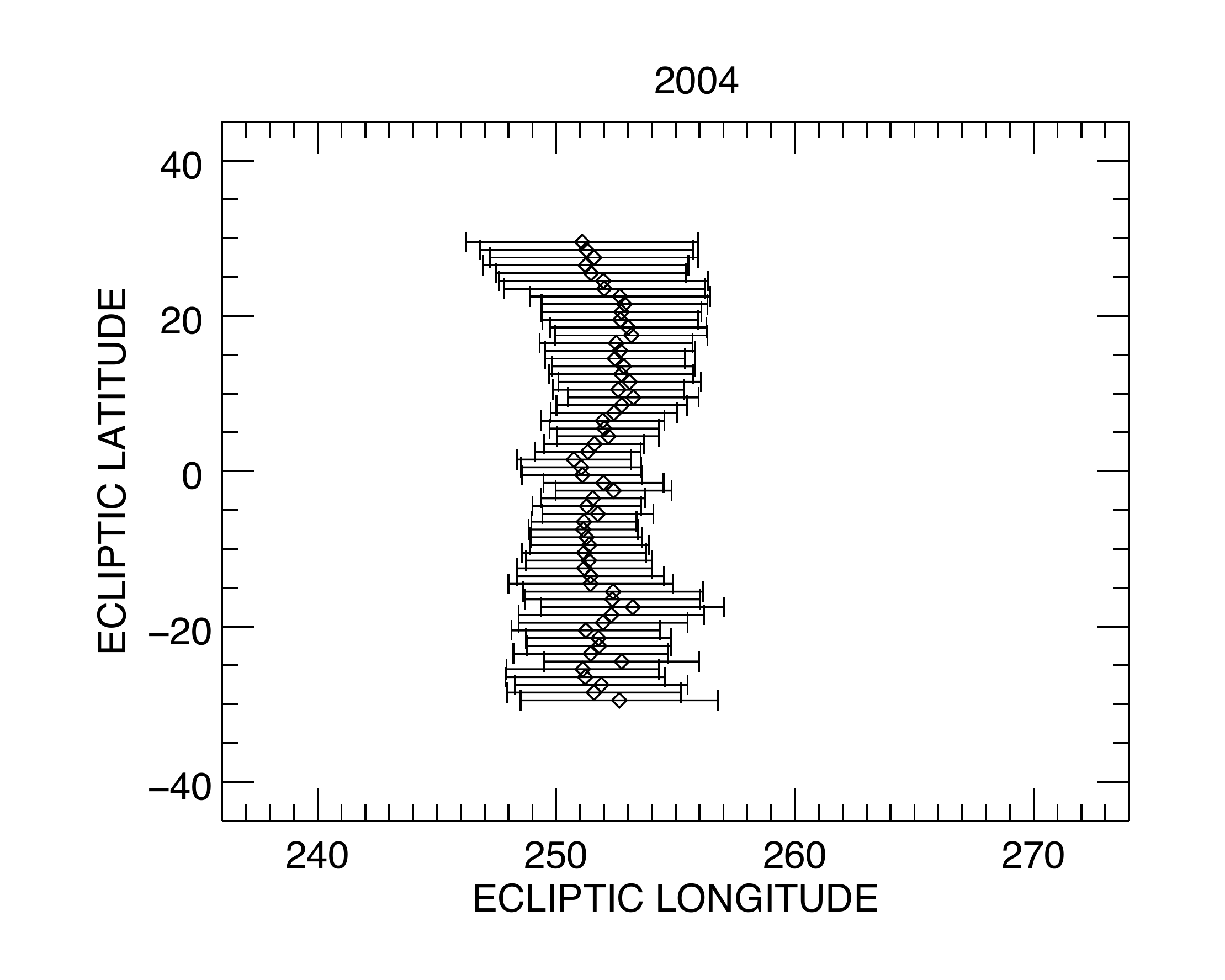}
   \includegraphics[scale=0.2]{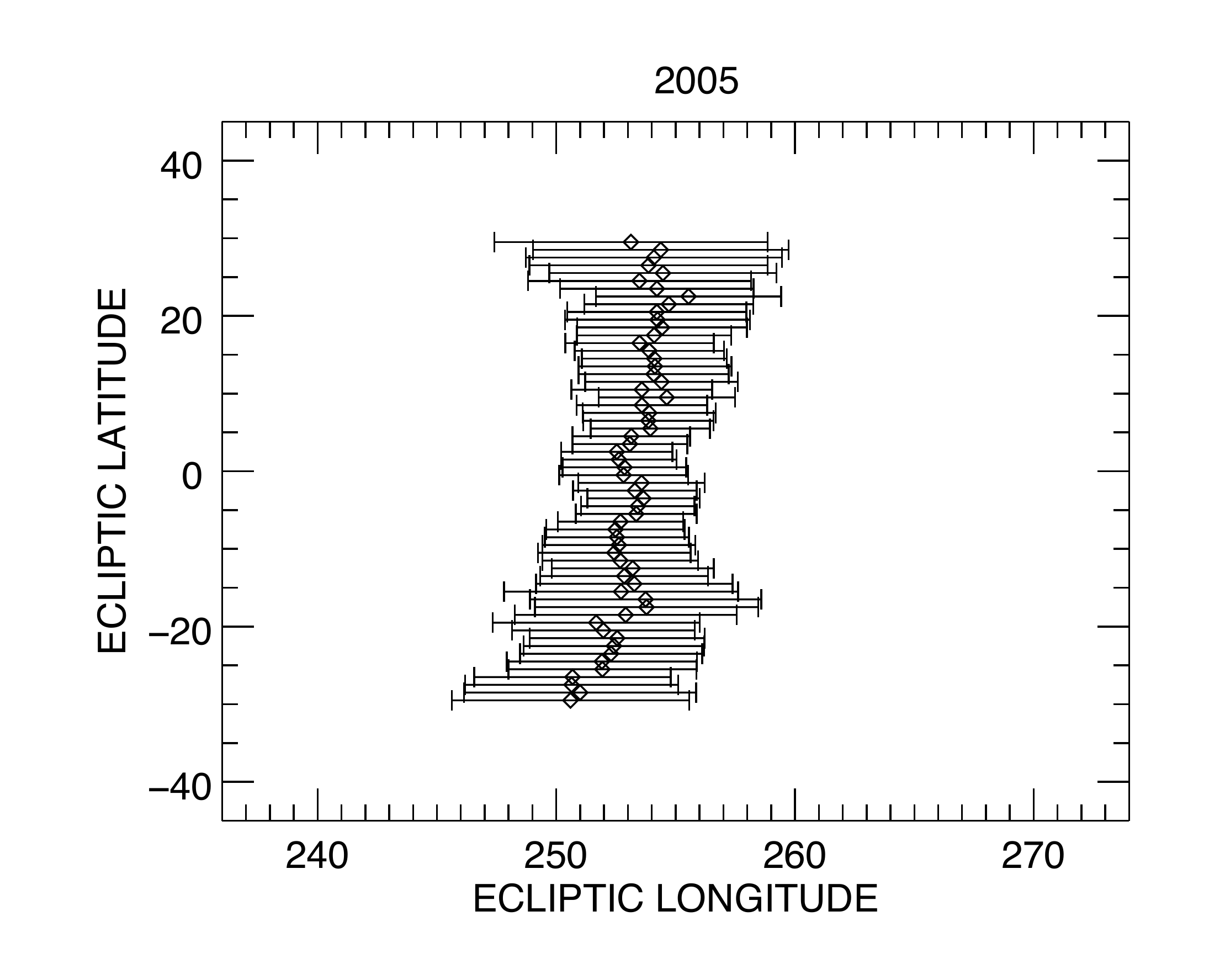}
   \includegraphics[scale=0.2]{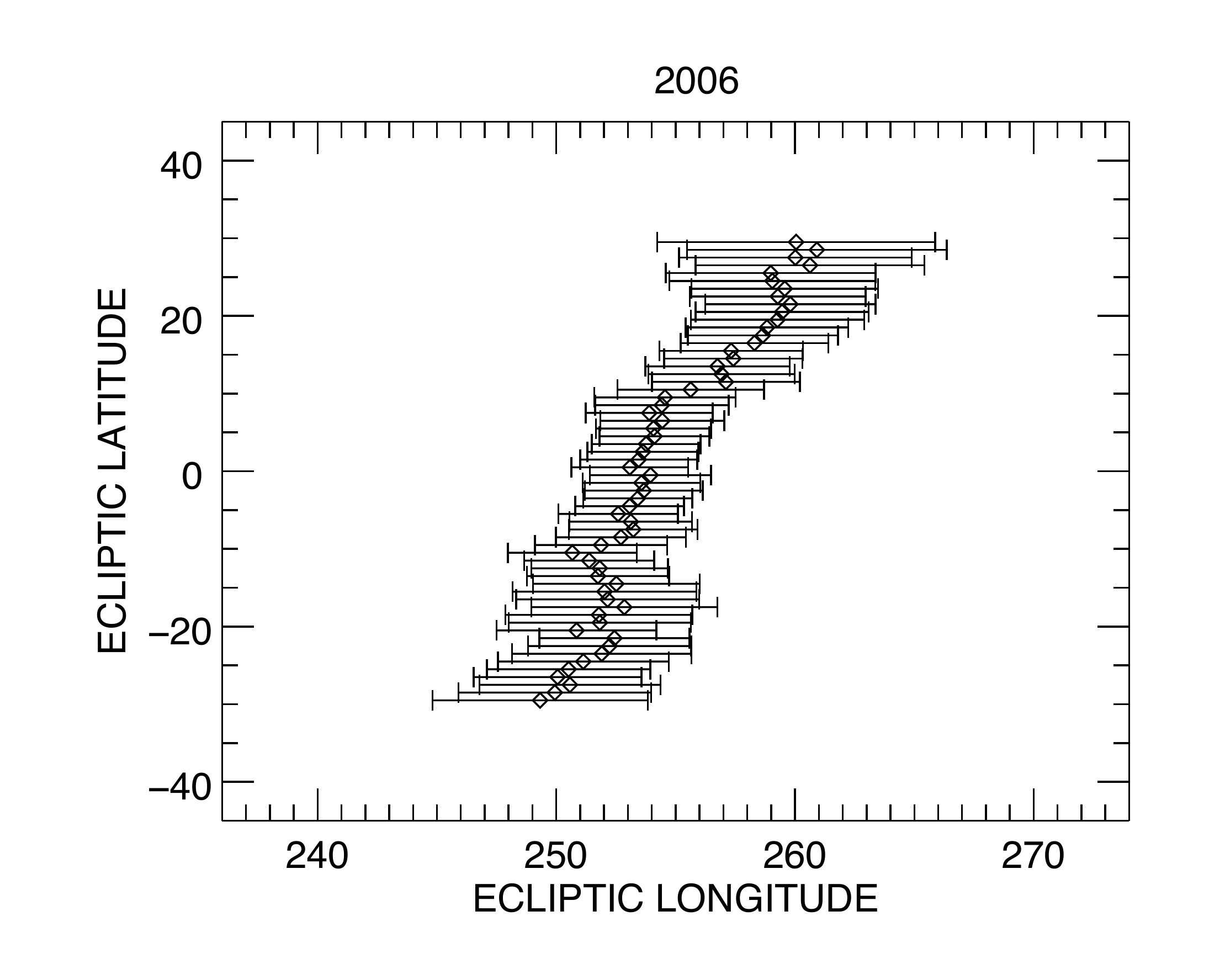}
   \includegraphics[scale=0.2]{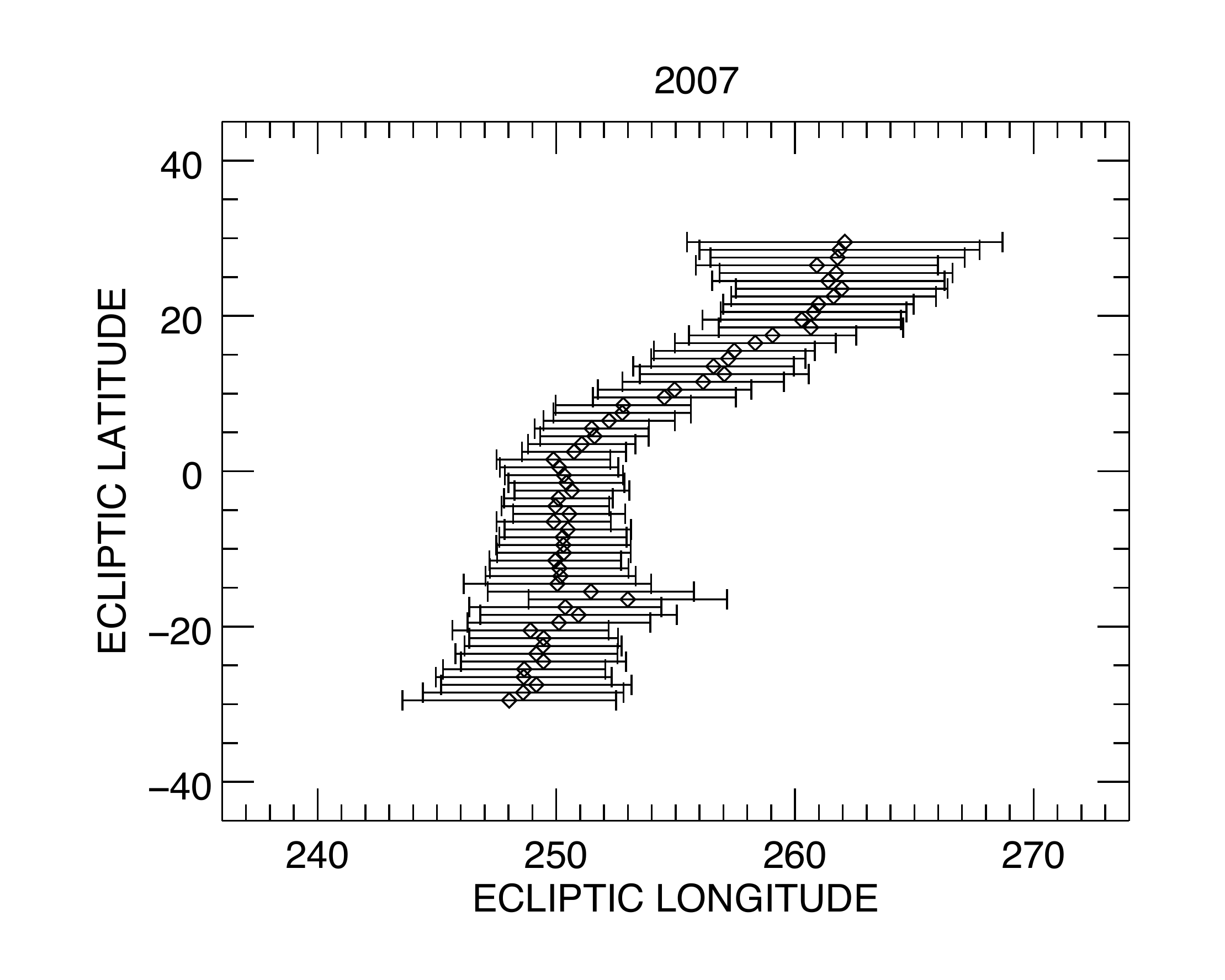}
   \includegraphics[scale=0.2]{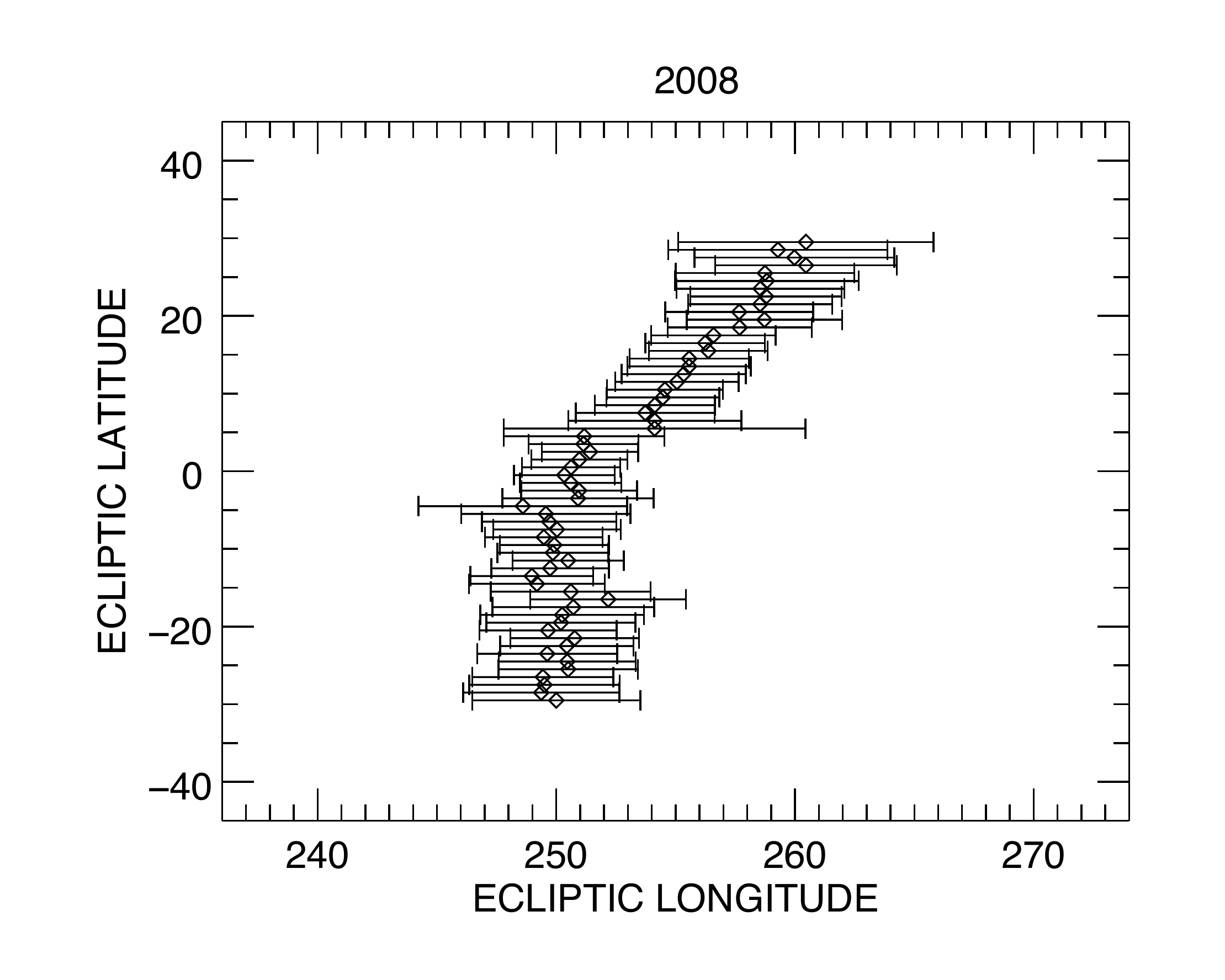}
   \includegraphics[scale=0.2]{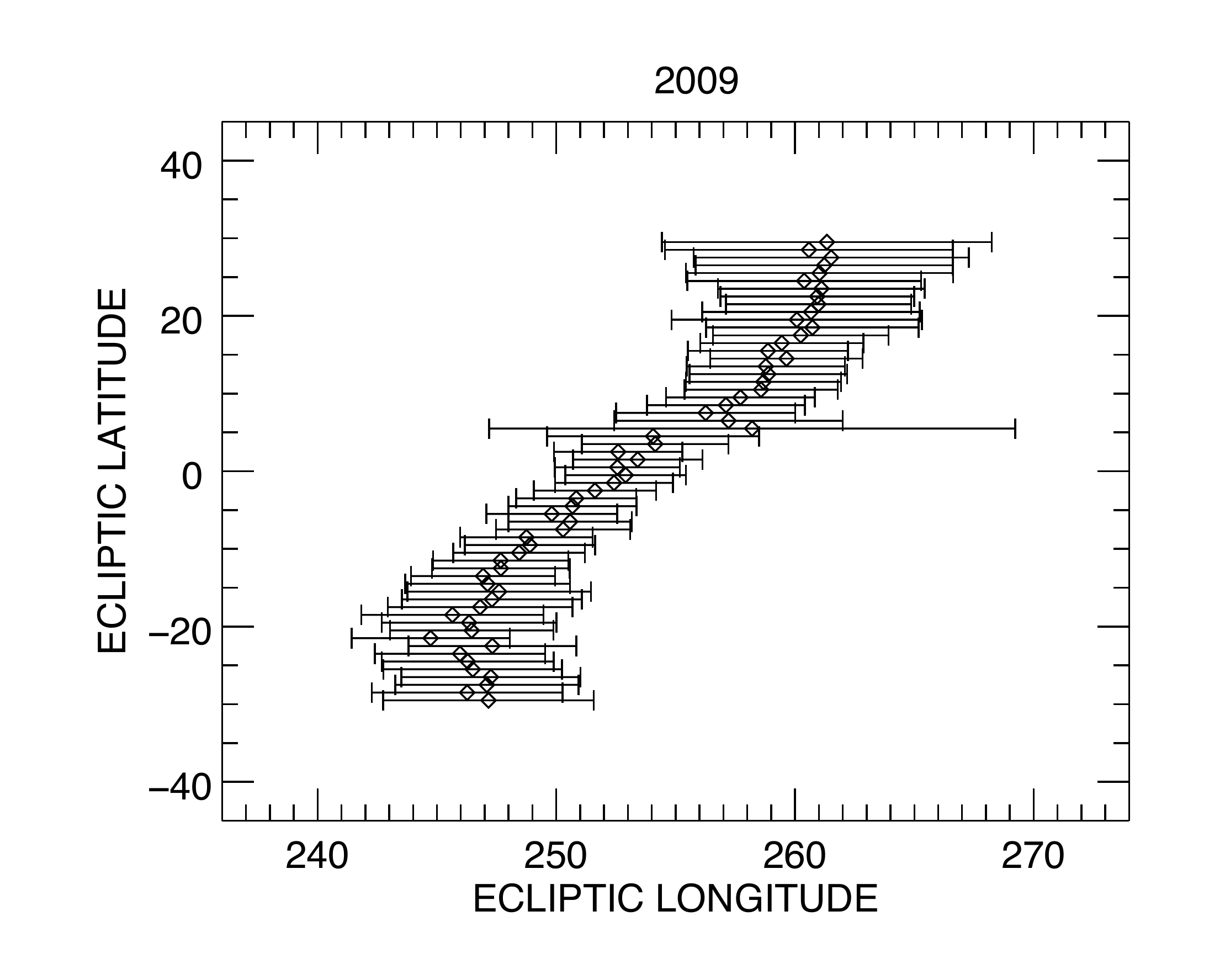}
   \includegraphics[scale=0.2]{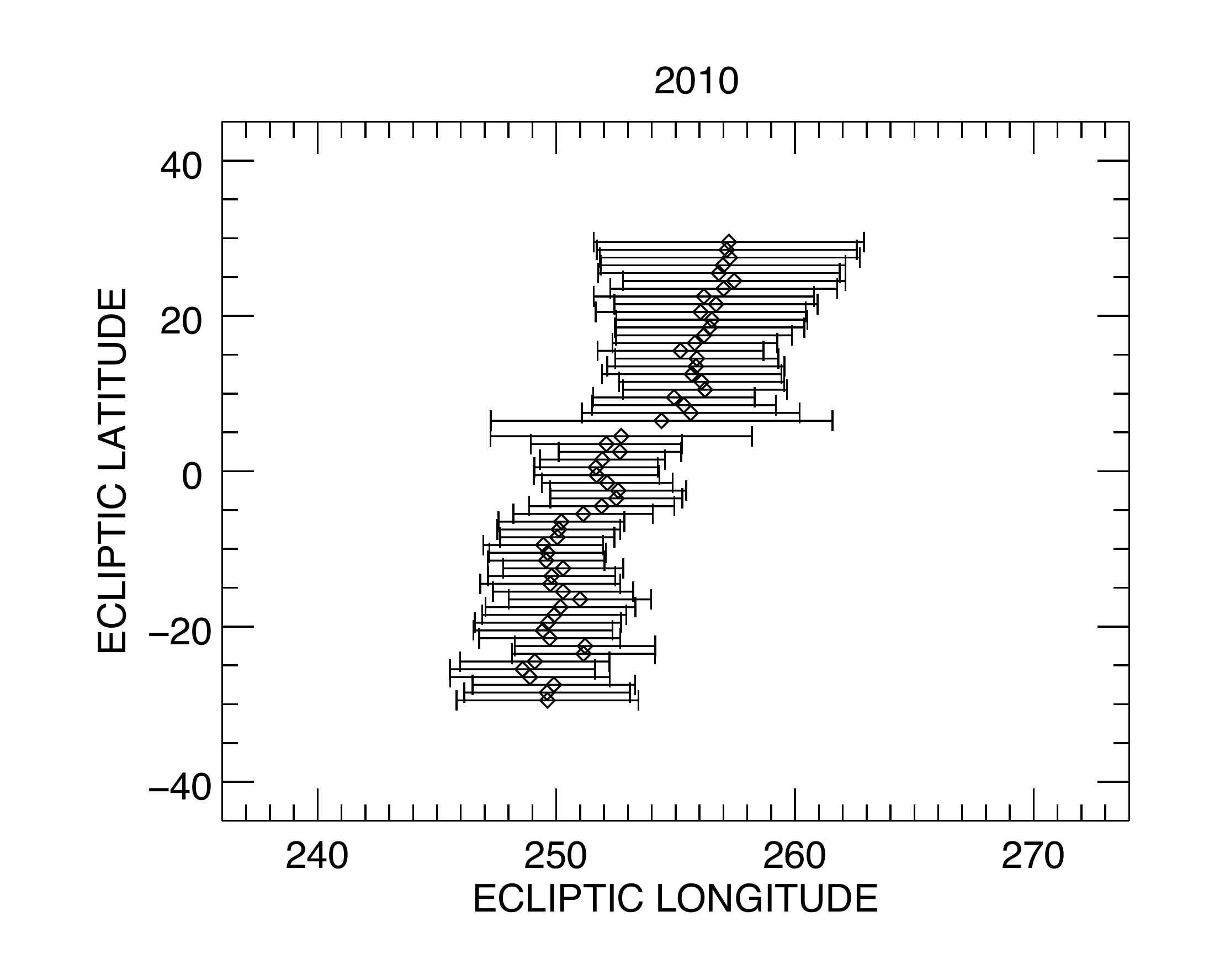}
   \includegraphics[scale=0.2]{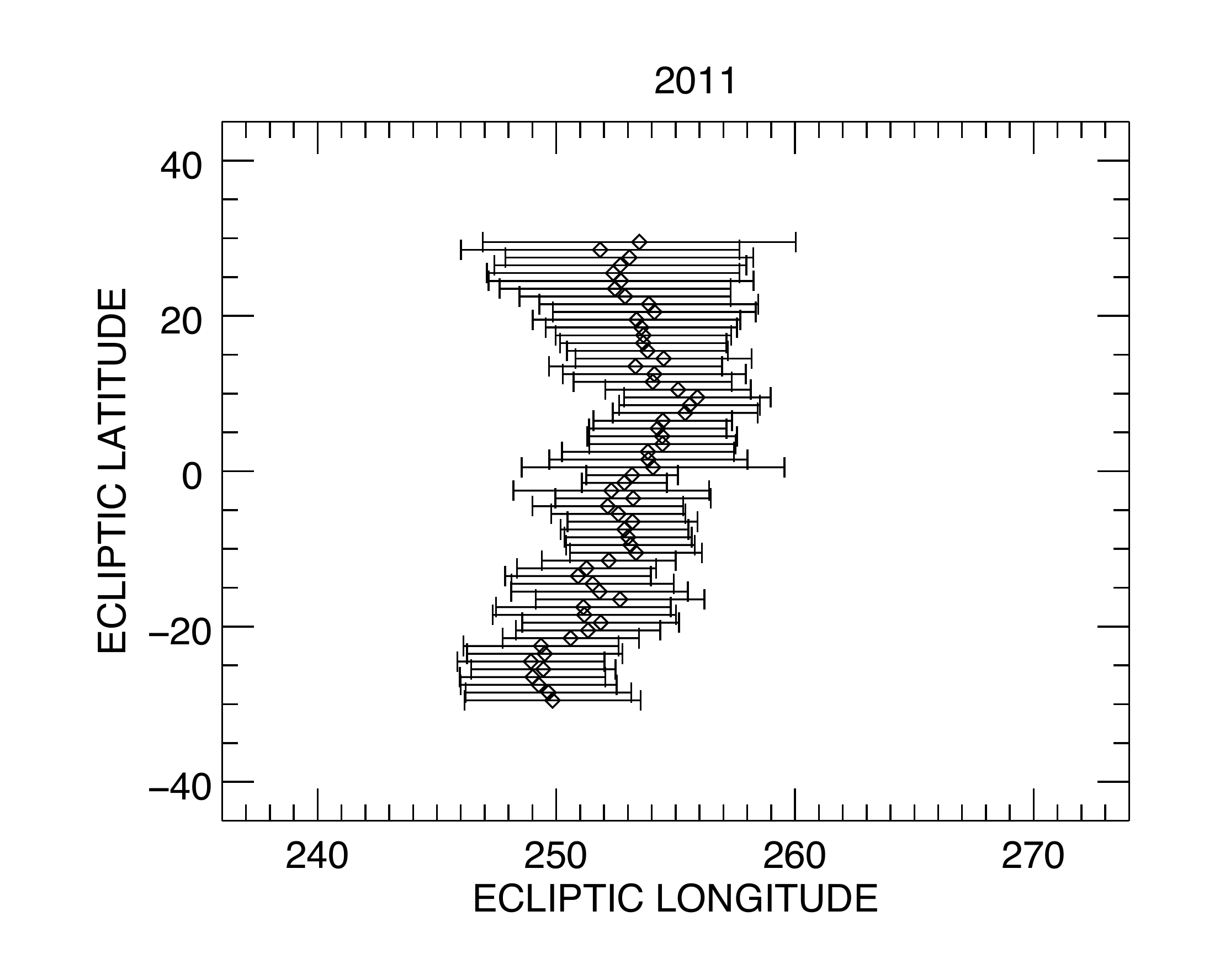}
   \includegraphics[scale=0.2]{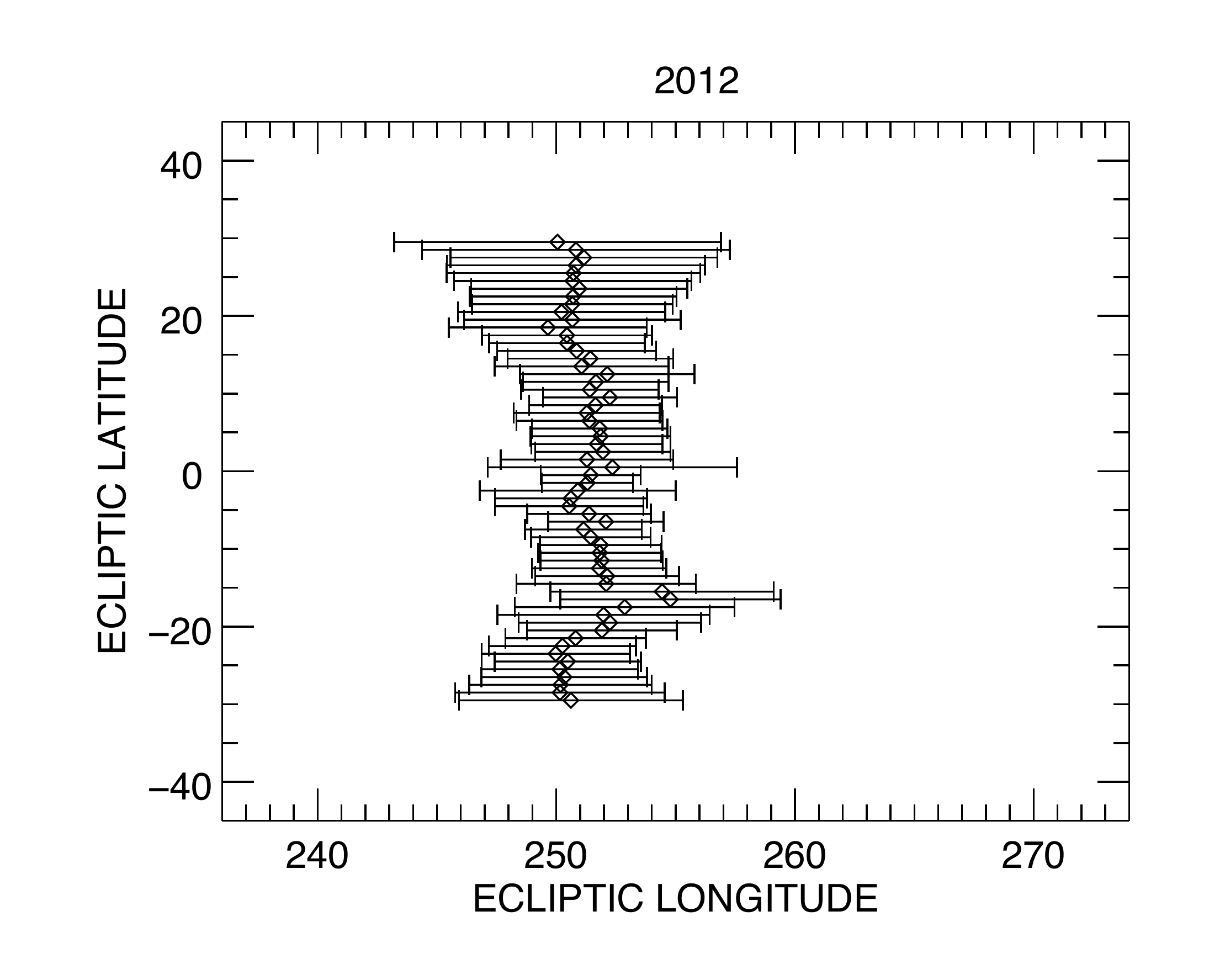}
   \includegraphics[scale=0.2]{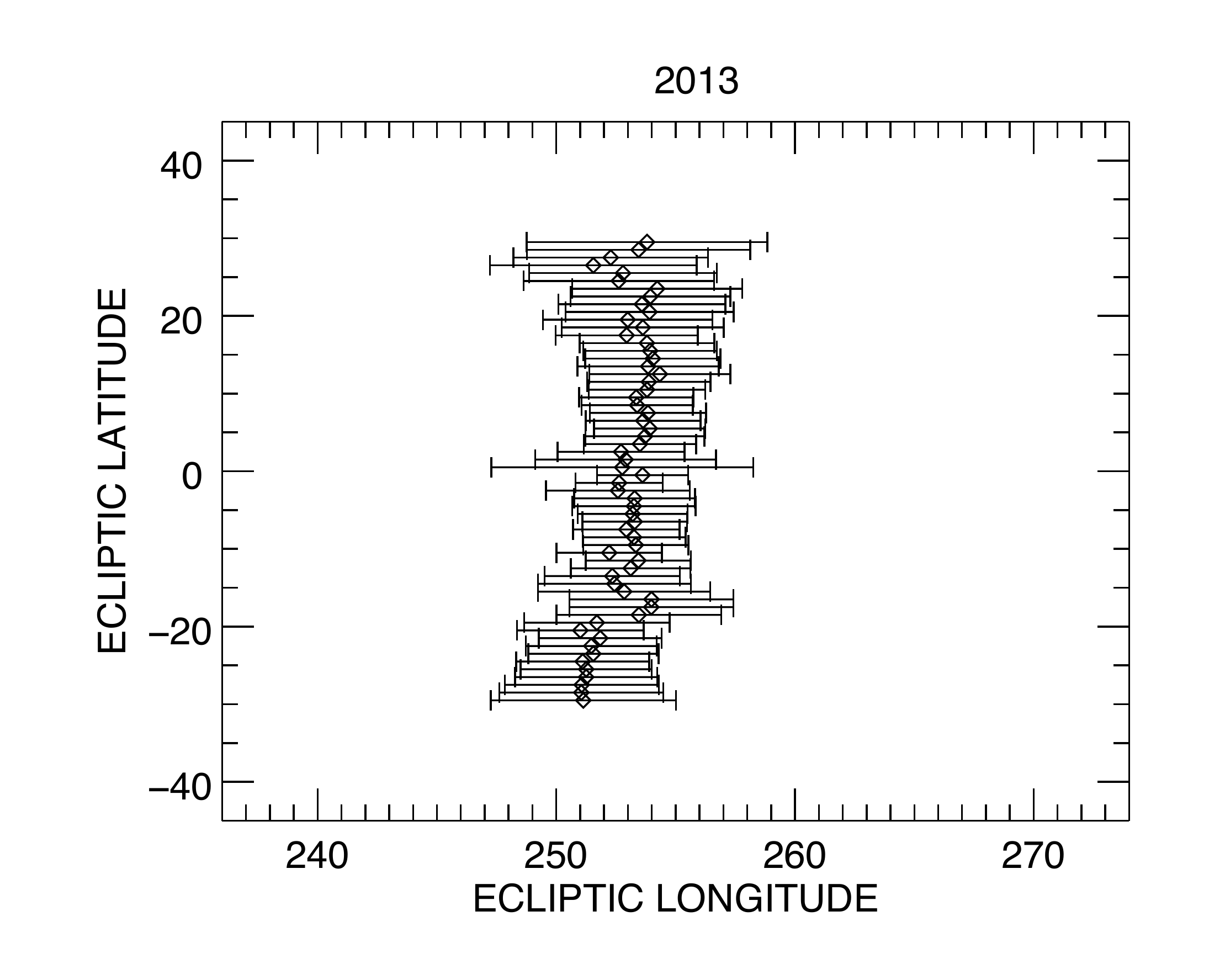}
   \includegraphics[scale=0.2]{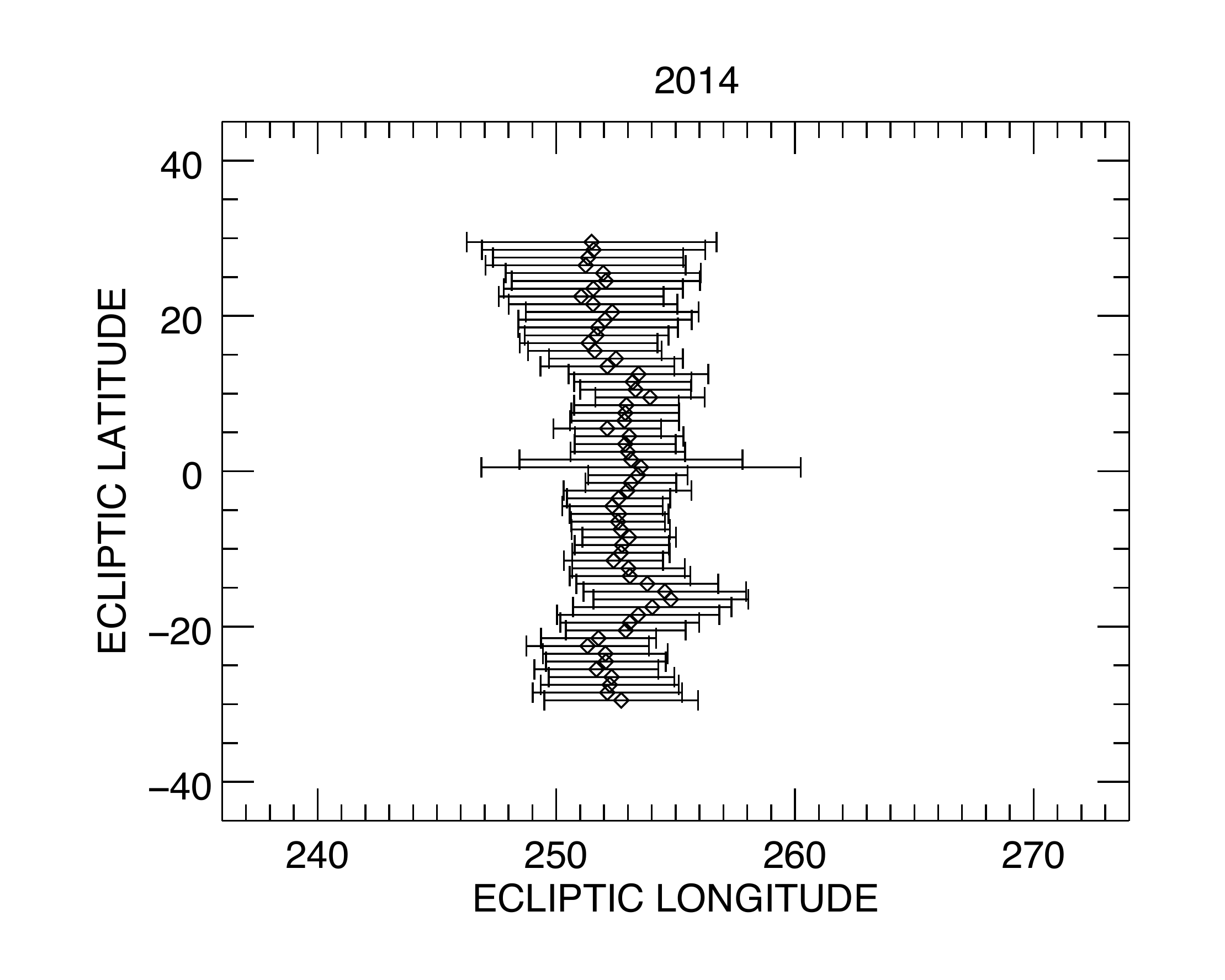}
   \includegraphics[scale=0.2]{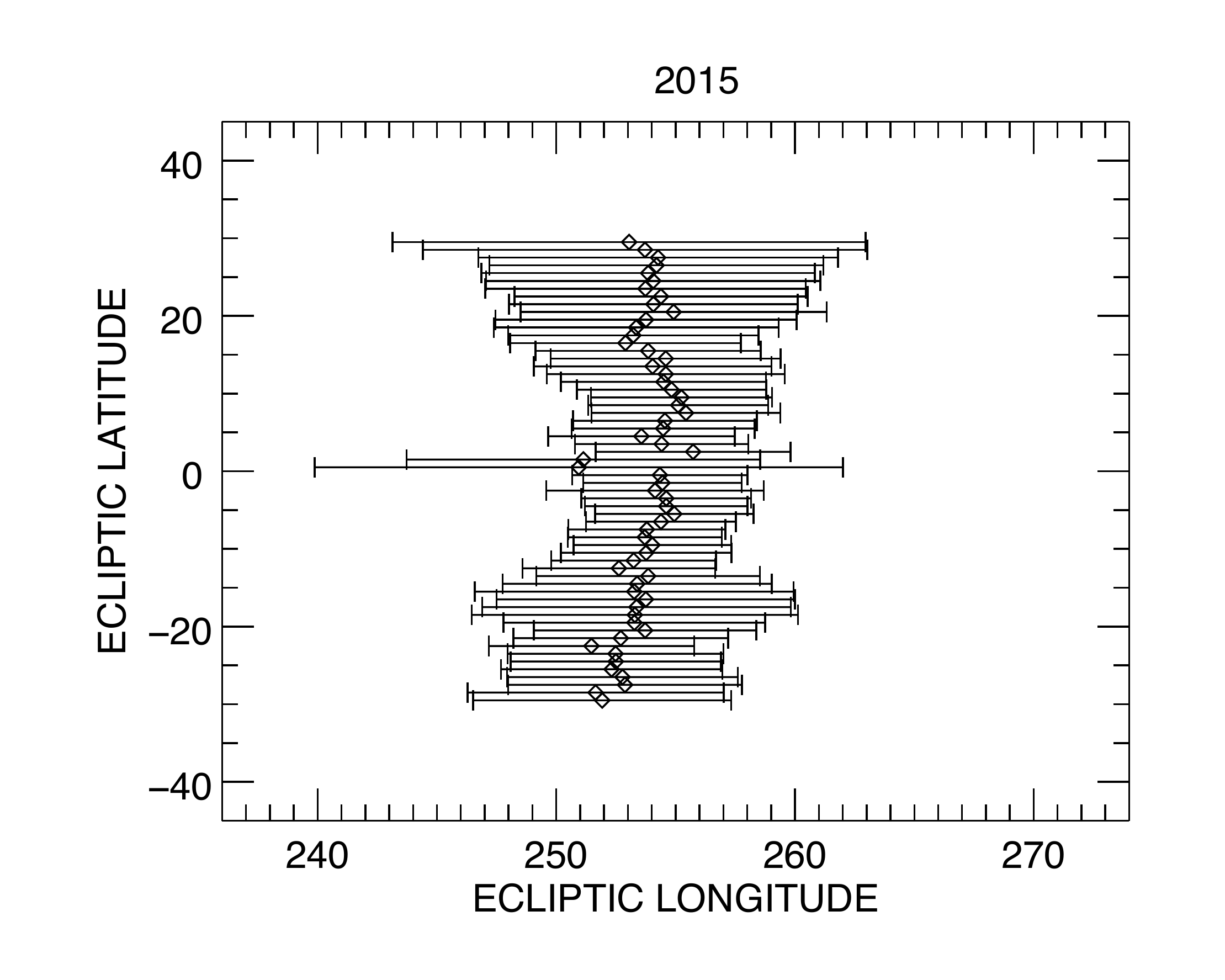}
   \caption{Results of the flow ecliptic longitude (X-axis) derived from fitting SWAN data over a $\pm$30\dg\ latitude range of coupled forward/backward pointing lines of sight for every year between 1996 and 2015.}
               \label{figLongErs}%
    \end{figure*}

Figure~\ref{figMaps} shows typical maps of the Lyman-\al\ background measured by SWAN during the minimum and maximum phase of the solar activity, composed of the two halves measured by the two sensors. Two areas in the sky cannot be observed: the largest is blocked by the spacecraft, while the smaller area corresponds to the shadows of the shields protecting the two sensors from direct sunlight.  Save for contamination from bright early-type stars (mainly in the southern ecliptic hemisphere), which is removed with a mask during data processing, the entire signal in this map is the solar Lyman-\al\ radiation backscattered by interstellar H atoms that are permanently flowing within the heliosphere.

The SWAN maps present a characteristic pattern reflecting the effect of solar activity (gravity/radiation pressure equilibrium on the atom trajectories), ionization processes, and Lyman-\al\ illumination on interstellar H atoms. In the upwind side of the flow ($\sim$253\dg\ ecliptic longitude), where H atoms approach closest to the Sun before they are ionized, the MER is observed, while on the downwind side ($\sim$73\dg\ ecliptic longitude) where H atoms are less likely to survive ionization, a so-called ionization cavity forms, which is devoid of neutral atoms. During solar minimum, when global ionization is lower, H atoms survive closer to the Sun, and the MER is located closer to the observer \citep{Lallement1984, Lallement1985}. Moreover, the higher ionization rate associated with the slow solar wind confined in the low heliographic latitude zone during solar minimum is responsible for the groove pattern observed in intensity maps, which divides the MER into two lobes roughly north and south of the ecliptic plane. This pattern had already been observed in the Prognoz 5 data \citep{Bertaux1996}, and has since been confirmed in solar minimum maps recorded by SWAN \citep[e.g., Fig.~\ref{figMaps} - top, and also in][]{Bertaux1997}. In reverse, during solar maximum, when the ionization rate becomes isotropic at most latitudes \citep{Quemerais2006a}, the groove tends to disappear and the MER shape becomes more circular-like and is centered only a few degrees north of the ecliptic plane (Fig.~\ref{figMaps} - bottom).

Owing to the MER and the ionization cavity (Fig.~\ref{figGeo}), the parallax effects, which are well documented in previous analyses of the SWAN data \citep{Quemerais2006a}, induce a modulation of the intensity that is observed for a given constant line of sight as SOHO orbits around the Sun. In general, the Lyman-\al\ intensity will increase when SOHO approaches the MER, while it will decrease when the spacecraft travels downwind through the ionization cavity. Since the interstellar H flow is, to first order, axisymmetric, it is possible to find viewing directions that scan through equal volumes of emitting interstellar H gas and therefore measure the same Lyman-\al\ intensity. When we assume that solar effects on the interstellar H distribution (Lyman-\al\ solar illumination and SW ionization processes) are averaged over solar rotation, these lines of sight should be symmetrical with respect to the plane, including the flow axis and the solar rotation axis. Then, the `equal emitting volume' condition is satisfied when the SOHO orbit intercepts this plane, twice a year on the upwind and downwind directions of the interstellar H flow axis.

Here, we make use of these parallax effects to determine the interstellar flow ecliptic longitude. We chose two directions that are symmetrical with respect to the plane including the SOHO-Sun line and the north ecliptic pole, instead of the solar rotation axis. Since the SOHO orbit around the Sun lies close to the ecliptic and the ascending node of the heliographic equator ($\lambda$ $\sim$ 75\dg) almost coincides with the interstellar axis downwind direction, it is simpler to assume the north ecliptic pole to define this symmetry plane. Then, as this plane rotates along with SOHO around the Sun, the `equal-emitting volume' condition is fulfilled when the plane is aligned with the interstellar flow axis, allowing us to determine the ecliptic longitude of the interstellar flow vector. 

For simplicity, for every SWAN star-corrected Lyman-\al\ intensity map we selected a line of sight perpendicular to the SOHO-Sun line at every 1\dg\ latitude between $\pm$30\dg\ from the ecliptic plane. In Fig.~\ref{figLya2time} we plot the Lyman-\al\ intensity for the coupled lines of sight at 0\dg\ latitude as a function of time. The annual modulation that is due to the motion of SOHO around the Sun is clearly visible, as well as the intersection of the curves when the SOHO-Sun line is aligned with the interstellar flow plane of symmetry and the two lines of sight point through an equal-emitting gas volume. The gaps in the data correspond to maps where the selected lines of sight points are removed because of starlight contamination. The effect of the 11-year solar activity cycle is also visible, with the solar maximum marked by the enhancement of the absolute Lyman-\al\ intensity in 2000-2002 (and to a lesser extent in 2011-2013). The annual trend of the data is the same in every same-latitude sample.

%-------------------------------------- Two column figure (place early!)
   \begin{figure}
   \centering
   \includegraphics[width=9cm]{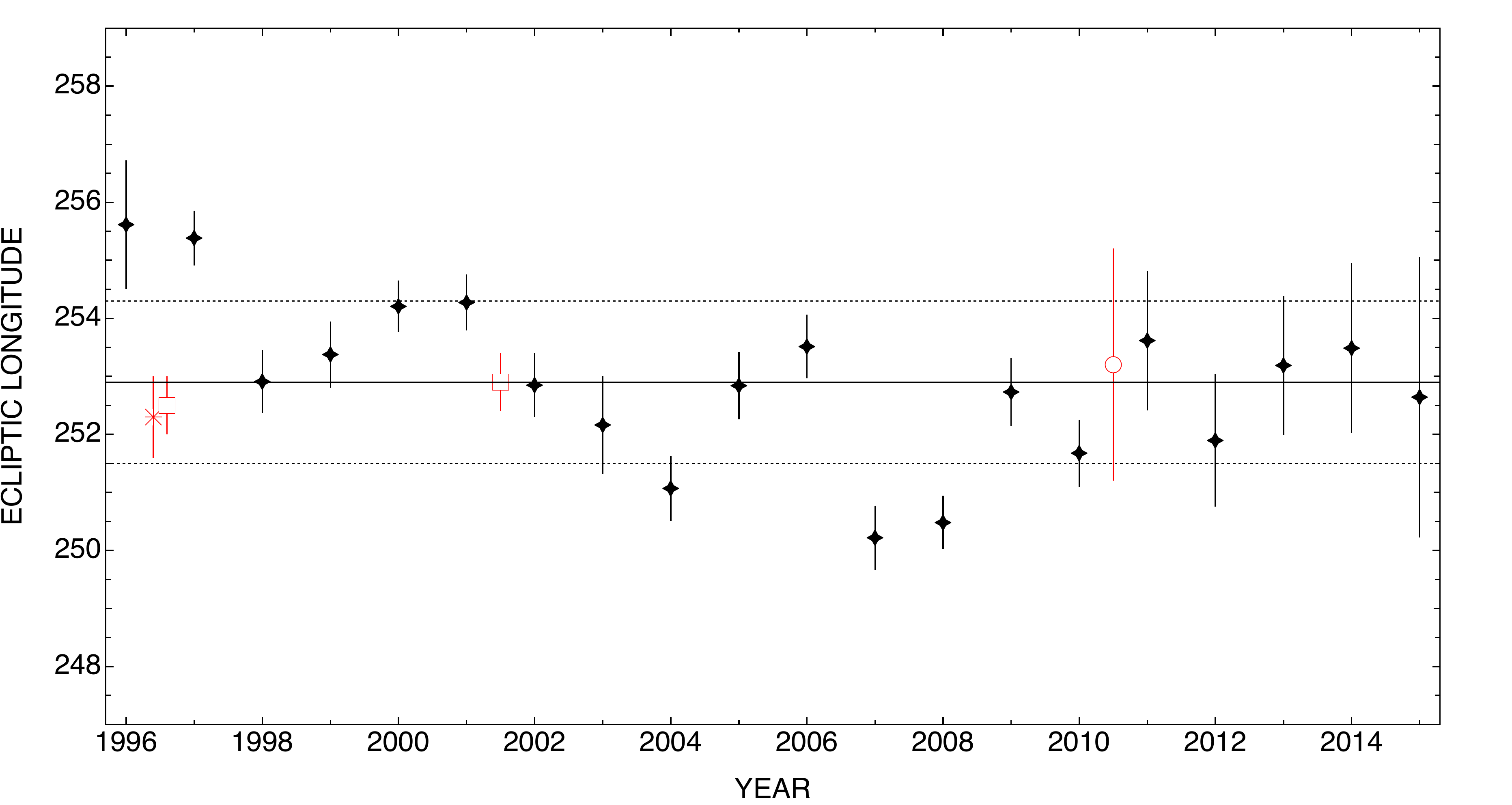}
   \caption{Annual results for the flow ecliptic longitude derived from the 0\dg\ latitude lines of sight. The horizontal full line represents the average value of the results in the ecliptic. Four previous determinations of the flow ecliptic longitude derived with different methods are also marked: the \cite{Quemerais1999} Lyman-$\alpha$ line profile reconstruction and projected velocity map based on 1996-1997 SWAN hydrogen-cell data (asterisk), the
\cite{Lallement2005,Lallement2010a} zero-Doppler-shift method from the same data in 1996 and 2001, respectively (squares), and the \cite{Quemerais2014} triangulation method from SWAN and MESSENGER coupled observations in 2010 (circle). All previous determinations lie within the standard deviation range (dotted lines) determined in our analysis.}
               \label{figLat0}%
    \end{figure}

We then extracted the annual Lyman-\al\ intensity for every same-latitude line of sight pair and plotted it as a function of the spacecraft ecliptic longitude. We corrected for the solar Lyman-\al\ variations by dividing by the solar Lyman-\al\ flux \citep{Woods2000} measured four days before the SWAN data for the lines of sight pointing forward, and four days after the SWAN data for the lines of sight pointing backward. As explained above, for symmetry reasons, the intensities should be equal when SOHO is along the wind axis, either upwind or downwind. We calculated the difference of the two coupled curves (forward - backward), and linearly fit the data as function of the SOHO ecliptic longitude. A fit example for the upwind position of SOHO is shown in Fig.~\ref{fig1999Fit}, where the smoothing effect of the solar Lyman-\al\ correction is clearly visible in the data. In general, the solar Lyman-\al\ correction improves the linear fit $\chi^2$ by several factors. Using the fit parameters ($\alpha$, $\beta$) and their error bars ($\sigma_{\alpha}$, $\sigma_{\beta}$) calculated from the SWAN data standard deviations, we solved the equation y = $\alpha x + \beta$ for y=0, where y is the Lyman-\al\ intensity difference and x is the SOHO ecliptic longitude in degrees. For each same-latitude line of sight pair, the ascending and descending slope equations give us two solutions, one for the upwind and one for the downwind ecliptic longitude of SOHO. In the following sections we discuss the results only for the upwind longitude, but the downwind analysis, results, and conclusions are the same.

%-------------------------------------------------------------
%                                             Simple A&A Table
%-------------------------------------------------------------
%
%\begin{table}
%\caption{UW Ecliptic Longitude for Solar Maximum Years}             % title of Table
%\label{table:1}      % is used to refer this table in the text
%\centering                          % used for centering table
%\begin{tabular}{c c c c}        % centered columns (4 columns)
%\hline\hline                 % inserts double horizontal lines
%\multicolumn{2}{c}{ Solar Cycle 23}    & \multicolumn{2}{c}{ Solar Cycle 24}\\
%Year      & UW Ecl. Long. (\dg) &  Year & UW Ecl. Long. (\dg) \\    % table heading 
%\hline                        % inserts single horizontal line
%   1999 & 253.7 $\pm$ 0.4 &  2011   & 252.6 $\pm$ 0.5\\      % inserting body of the table
%   2000 & 253.7 $\pm$ 0.3 &  2012   & 251.3 $\pm$ 0.5\\
%   2001 & 253.4 $\pm$ 0.4 &  2013   & 252.8 $\pm$ 1.2\\
%   2002 & 252.9 $\pm$ 0.4 &  2014   & 252.5 $\pm$ 0.4\\
%   2003 & 252.6 $\pm$ 0.6 &  2015   & 253.8 $\pm$ 1.2\\ 
%   2004 & 251.9 $\pm$ 0.4 &     & \\ 
%   2005 & 253.2 $\pm$ 0.5 &     & \\ 
%\hline                                   %inserts single line
%\end{tabular}
%\end{table}

%-------------------------------------------------------------
%
\section{Discussion}\label{Sec3}

The annual results for the upwind ecliptic longitude are presented in Fig.~\ref{figLongErs}. The main prominent feature in the results is a modulation of the ecliptic longitude derived from high-latitude data, especially during the solar minimum periods (1996-1998 and 2006-2010). During solar minimum, when the MER is divided into two lobes by the Lyman-\al\ groove, the derived flow longitude shifts toward higher values in the northern hemisphere and toward lower values in the southern hemisphere. On the other hand, at 0\dg\ latitude, the flow upwind ecliptic longitude value is stable throughout the solar cycle at an average value of 252.9\dg$\pm$1.4\dg\ (Fig.~\ref{figLat0}). In the same figure we include previous determinations from i) a model-independent method of Lyman-$\alpha$ line profile reconstruction and projected velocity map based on 1996-1997 SWAN hydrogen-cell data \citep[252.3\dg $\pm$ 0.7\dg;][]{Quemerais1999}, ii) the zero-Doppler-shift localization method, fitting a hot model to the same data in 1996-1997 \citep[252.5\dg $\pm$ 0.5\dg;][]{Lallement2005} and 2001-2002 \citep[252.9\dg $\pm$ 0.5\dg;][]{Lallement2010a}, and iii) the MER triangulation method from SWAN and MESSENGER coupled observations in 2010 \citep[253.2\dg $\pm$ 2\dg;][]{Quemerais2014}. All of these previous determinations lie within the standard deviation range (dotted lines in Fig.~\ref{figLat0}) derived in our analysis. As seen in Fig.~\ref{figLongErs}, for solar maximum periods (1999-2005 and 2011-2015), the upwind ecliptic longitude values for all latitudes tend to converge toward the same value as the one derived for lines of sight near the ecliptic, consistent with the quasi-circular and well-centered shape of the MER near the ecliptic plane. 

\begin{figure*}
   \centering
   \includegraphics[scale=0.3]{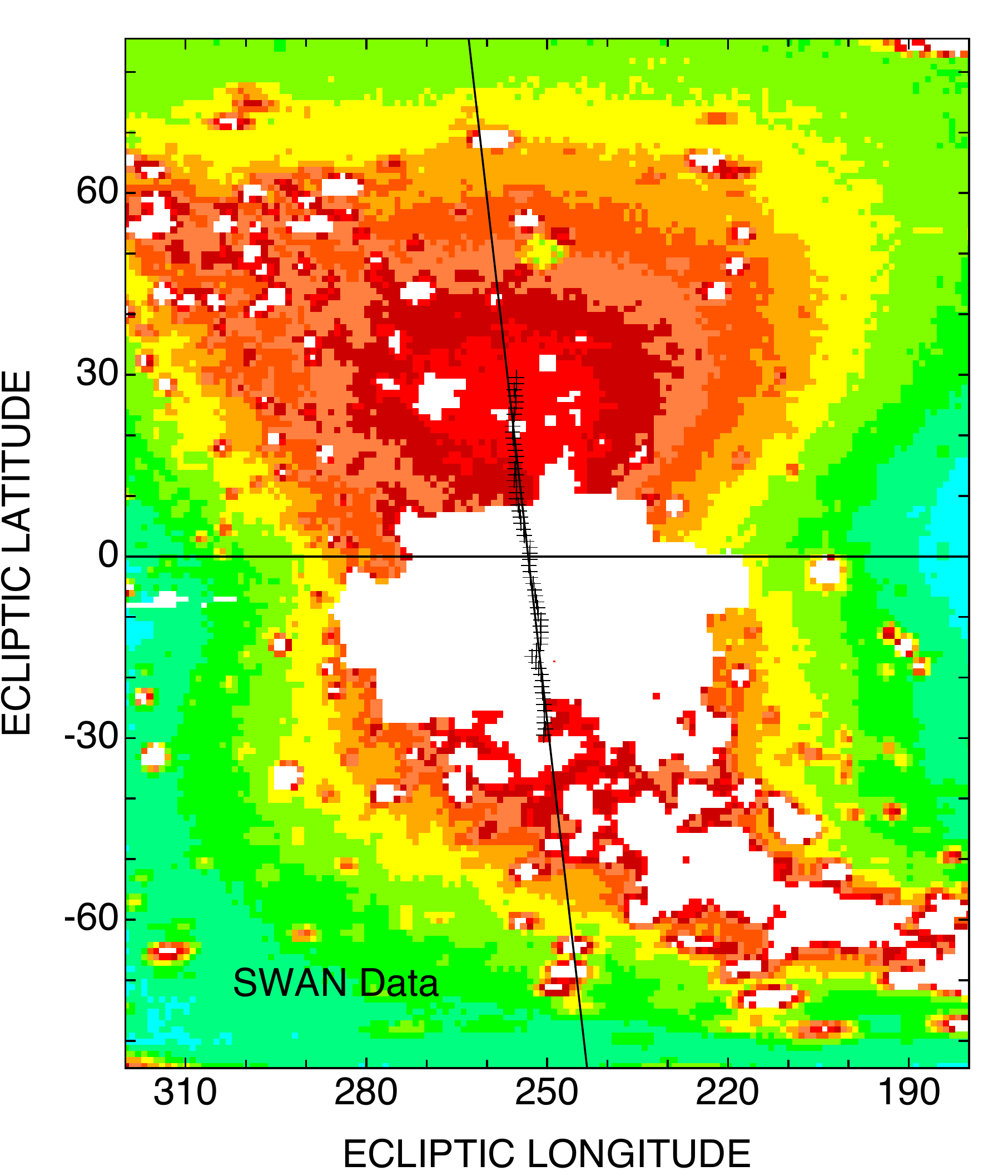}
   \includegraphics[scale=0.3]{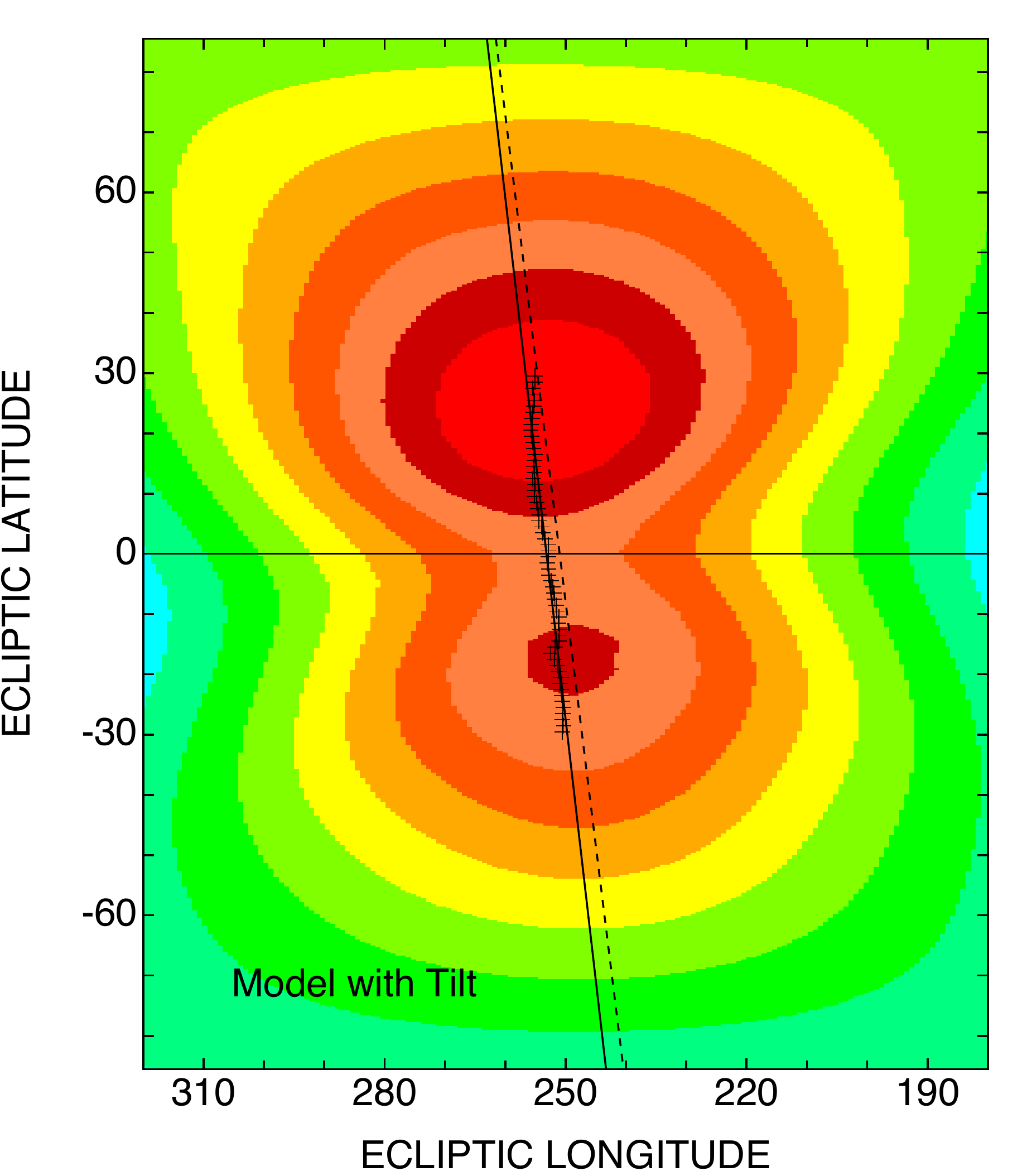}
   \includegraphics[scale=0.3]{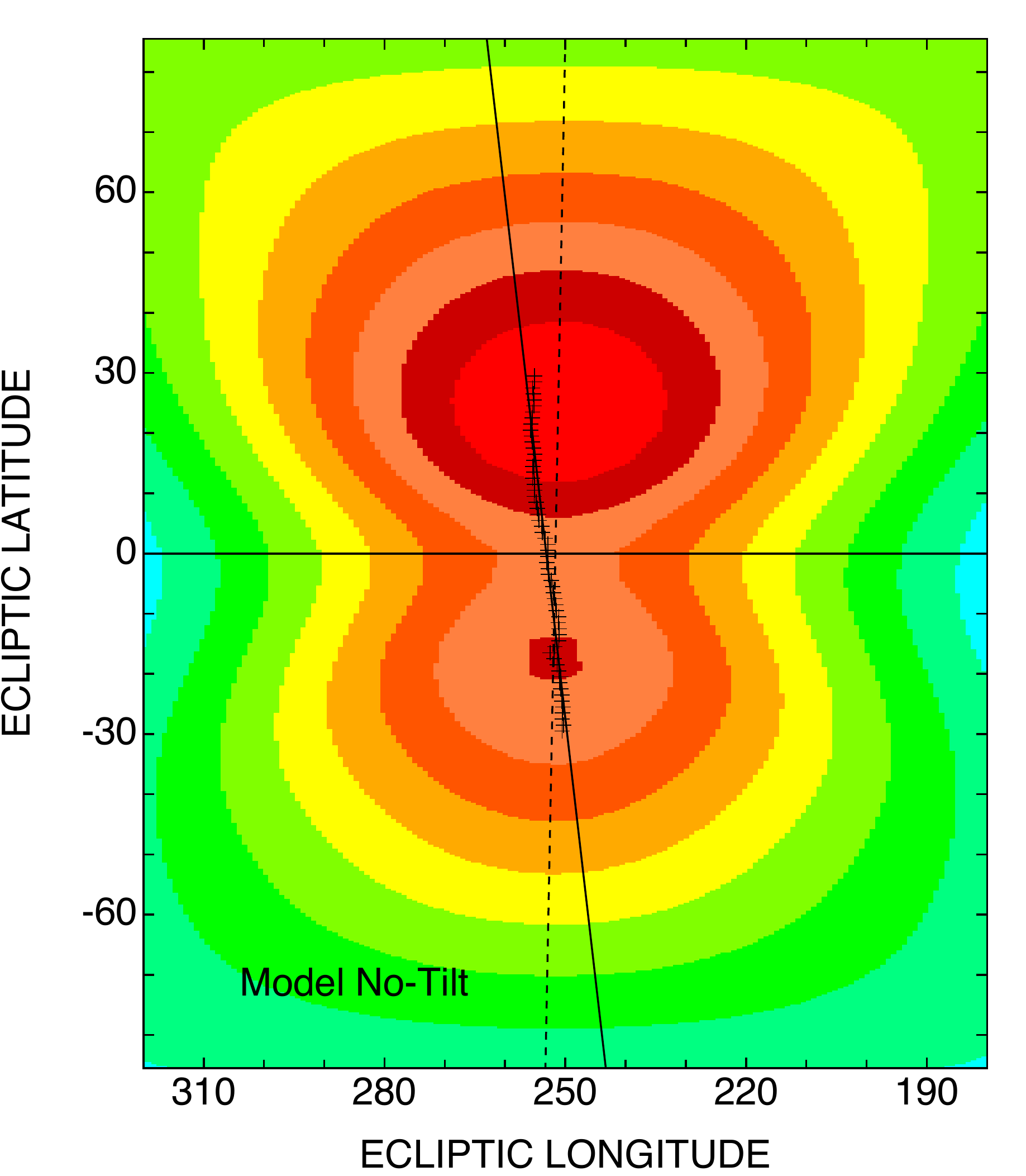}
   \caption{Lyman-$\alpha$ maps of the upwind region, showing the MER and groove, as seen by an observer on the downwind position for solar minimum. We show the SWAN data for Dec. 3, 2009 (left), a simulation result calculated for this date using the Moscow full-heliospheric model (middle), and the same simulation assuming that the solar rotation axis is perfectly aligned with the ecliptic axis (right). The black crosses represent the average upwind longitude derived from this analysis for the different latitude parallels. The black line approximately traces the north-south direction on which these directions are aligned, while the dashed lines in the center and right panels show the alignment of the MER north and south lobes in the model maps. No such alignment could be precisely defined from the data maps, as there was a large gap in the data that is due to the detector shield. }
               \label{figGroove}%
    \end{figure*}

The reason for the high-latitude modulation lies in the shape and inclination of the MER off-ecliptic lobes. In our method we assumed that the interstellar flow is perfectly axisymmetric at infinity and that the interstellar plane of symmetry includes the ecliptic pole axis. However, the lobes are not exactly aligned with the ecliptic pole axis, but rather with the solar rotation axis, since the effects of Lyman-\al\ illumination, lower radiation pressure during solar minimum, and charge-exchange ionization processes are averaged in the SWAN maps by solar rotation. To demonstrate this effect, we used two simulations of the heliosphere during solar minimum conditions performed with the IKI global time-dependent 3D model \citep{Izmodenov2013}. In the first case, the solar rotation axis is tilted to its natural value of 7.25\dg, while in the second, it was considered with no tilt at all with respect to the ecliptic pole axis. In Fig.~\ref{figGroove} we present the results of the simulations on the shape of the MER during solar minimum, along with a SWAN data projection of the MER region and the average upwind longitudes derived from all latitude parallels. Despite the simplified assumptions in our method, the effect is undeniable, since the figure shows that the MER lobes and the averaged flow longitudes derived from this analysis align when the solar rotation axis is tilted.

%-------------------------------------- Two column figure (place early!)
%

%-----------------------------------------------------------------

\section{Conclusions}
We have used SWAN interplanetary Lyman-\al\ data and the parallax effects induced by the SOHO annual motion in the Lyman-\al\ intensity to calculate the ecliptic longitude at which the spacecraft crosses the interstellar H-flow plane of symmetry and thus derived the ecliptic longitude of the interstellar H-flow vector. In the last 20 years of SWAN operations we find no significant change in the interstellar H-flow vector longitude, except for a modulation at higher latitudes that is correlated with the solar cycle. This modulation is mainly due to the presence of the Lyman-\al\ groove pattern in the MER produced by the highly anisotropic SW during solar minimum. Near the ecliptic plane, where variations are considerably reduced, the average value for the flow upwind ecliptic longitude is found to be 252.9\dg\ $\pm$ 1.4\dg, in very good agreement with previous determinations \citep{Quemerais1999,Lallement2005,Lallement2010a,Quemerais2014}. In the years of solar maximum, the results at all latitudes tend to converge toward this `canonical' value, supporting the general idea of a more isotropic SW and a quasi-circular MER centered near the ecliptic plane. These results further support the argument in favor of the long-term stability of the interstellar vector for the H flow and in consequence for He as well. A more detailed analysis of the SWAN H-absorption cell data should allow a firm determination of the interstellar H flow vector in both longitude and latitude ($\lambda$, $\beta$) throughout the last solar cycles.

\begin{acknowledgements}
SOHO is a mission of international cooperation between ESA and NASA. We are grateful to ISSI for their support of the working team 223, through which this analysis has been developed.  We also wish to thank the French space agency CNES for their continued support of the SWAN instrument through the program `Soleil H\'eliosph\`ere Magn\'etosph\`ere'. OK is supported by RFBR grant No. 16-52-16008-CNRS-a.
\end{acknowledgements}

   \bibliographystyle{aa} % style aa.bst
   \bibliography{LyaCrossBib} % your references Yourfile.bib

%\begin{thebibliography}{}

%\end{thebibliography}

\end{document}